\documentclass[12pt,a4paper]{article}

\usepackage{epsfig}
\usepackage{amsmath,amsfonts,amssymb}
\usepackage{t1enc}
\usepackage{cite}

\providecommand{\openone}{\leavevmode\hbox{\small1\kern-3.8pt\normalsize1}}

\newcommand{\Dppmm}{\Delta^{\pm \pm}}

\newcommand{\Dpm}{\Delta^\pm}

\newcommand{\Esix}{\text{E}_6}
\newcommand{\Zlp}{Z'_\lambda}

\newcommand{\SM}{\Sigma_\text{M}}
\newcommand{\SD}{\Sigma_\text{D}}
\newcommand{\ENd}{NE_\text{d}}
\newcommand{\Es}{E_\text{s}}
\newcommand{\NM}{Z'N_\text{M}}
\newcommand{\ND}{Z'N_\text{D}}

\newcommand{\ptmiss}{p_T\!\!\!\!\!\!\!\!\not\,\,\,\,\,\,\,}
\newcommand{\tria}{{\tt Triada}}

\begin{document}

\begin{center}
\begin{Large}
{\bf Heavy lepton pair production at LHC: \\[2mm]
model discrimination with multi-lepton signals}
\end{Large}

\vspace{0.5cm}
J. A. Aguilar--Saavedra  \\[0.2cm] 
{\it Departamento de F\'{\i}sica Te\'orica y del Cosmos and CAFPE, \\
Universidad de Granada, E-18071 Granada, Spain} \\[0.1cm]
\end{center}

\begin{abstract}
If new leptons exist close to the electroweak scale, they can be produced in pairs at LHC through standard or new interactions. We study the production of heavy lepton pairs in SM extensions with: (i) a Majorana or Dirac lepton triplet, as those appearing in type-III seesaw; (ii) a lepton isodoublet $(N \, E)_{L,R}$; (iii) a charged isosinglet $E_{L,R}$; (iv) a Majorana or Dirac neutrino singlet $N$ and an additional $Z'$ gauge boson. It is shown that the trilepton final state $\ell^\pm \ell^\pm \ell^\mp$, which has a small SM background, constitutes the golden channel for heavy neutrino searches, being very sensitive to Majorana or Dirac neutrinos in triplet, doublet or singlet $\text{SU}(2)_L$ representations.
For higher luminosities, signals in this final state can also distinguish lepton triplets from doublets and singlets. The Majorana or Dirac nature of the heavy neutrinos is revealed by the presence or not of like-sign dilepton $\ell^\pm \ell^\pm$ signals without significant missing energy.
Notably, large $\ell^\pm \ell^\pm$ signals but with large missing energy are characteristic of Dirac triplets, distinguishing them from the other two models with a heavy Dirac neutrino.
Further discrimination is achieved with the analysis of the clean $\ell^+ \ell^+ \ell^- \ell^-$ final state.
\end{abstract}

\section{Introduction}
\label{sec:1}

Being a discovery machine, the Large Hadron Collider (LHC) will hopefully uncover any new physics close to the electroweak scale. In particular, the several variants of the seesaw mechanism 
proposed to explain light neutrino masses
can be tested~\cite{delAguila:2008cj,delAguila:2008hw}. In case that
one of these mechanisms is responsible for the light neutrino mass generation and seesaw messengers exist around the TeV scale or below, positive signals could be observed at LHC.

There are three types of tree-level seesaw which can originate light neutrino Majorana masses.
The original seesaw \cite{Minkowski:1977sc,GellMann:1980vs,Yanagida:1979as,Mohapatra:1979ia}, also known as seesaw I, introduces right-handed neutrino singlets $N$. Seesaw II \cite{Magg:1980ut,Cheng:1980qt,Gelmini:1980re,Lazarides:1980nt,Mohapatra:1980yp}
enlarges the SM with a complex scalar triplet $\Delta$ with hypercharge $Y=1$,
which contains three scalars $\Dppmm$, $\Dpm$, $\Delta^0$.
Seesaw III \cite{Foot:1988aq,Ma:1998dn} introduces lepton triplets $\Sigma$ with $Y=0$, each containing a charged lepton $E$ and a neutral one $N$.
In these three seesaw mechanisms, the lepton number violating (LNV) operator~\cite{Weinberg:1979sa}
\begin{equation}
(O_5)_{ij} = \frac{C_{ij}}{\Lambda} \overline{L_{iL}^c} \tilde \phi^* \tilde \phi^\dagger L_{jL} \,,
\label{ec:O5}
\end{equation}
is generated after integration of the heavy degrees of freedom,
where $L_{iL}$ are the SM lepton doublets with $i$ a flavour index, and $\phi$ is the SM Higgs. Subsequently, when electroweak symmetry is spontaneously broken, this operator gives light neutrino Majorana masses. The scale at which this operator is generated ({\em i.e.} the mass scale of the seesaw messengers) is not necessarily very high but it might happen that it is around the TeV. In this case,
the new leptons predicted in models of type-I and type-III seesaw might be produced at LHC.\footnote{In seesaw II the scalars predicted can also be produced
and observed in this mass range \cite{delAguila:2008cj} (see also
Refs.~\cite{Huitu:1996su,Gunion:1996pq,Akeroyd:2005gt,Hektor:2007uu,Han:2007bk,Perez:2008zc}).}
For seesaw III, if the new heavy states $E$, $N$ have masses up to several hundreds of GeV, they can be produced and observed at LHC already at the low luminosity phase~\cite{delAguila:2008cj}.
For seesaw I, the production of heavy neutrino singlets $N$ through SM interactions is very suppressed because of the small heavy neutrino mixing with SM particles, and signals are unobservable except for relatively ``light'' masses around 150 GeV~\cite{delAguila:2007em}. However, if new interactions exist, either $W'$ bosons~\cite{Keung:1983uu,Datta:1992qw,Ferrari:2000sp,Gninenko:2006br,delAguila:2009bb}, $Z'$ bosons~\cite{delAguila:2007ua,Huitu:2008gf,Blanchet:2009bu,Ma:2009gu} or new scalars \cite{Ma:2000cc,BarShalom:2008gt}, heavy neutrino singlets can be copiously produced, either singly or in pairs depending on the model.

The observability of the seesaw I--III signals and the discrimination among these models has been investigated in Ref.~\cite{delAguila:2008cj}. There,
an exhaustive analysis of all final states was carried out with a complete signal and background calculation, and the characteristic features of each seesaw model were highlighted. In this paper we will perform a complementary analysis. Our main objective here is to identify the relevant signals whose observation or non-observation would discriminate among different models with new heavy leptons. In particular, we want to design a strategy to determine if new leptons eventually observed at LHC could mediate a type-I or type-III seesaw mechanism. We will study heavy lepton pair production
\begin{align}
& pp \to E^+ E^- \,, \notag \\
& pp \to E^\pm N \,,\notag \\
& pp \to N N \,,
\end{align}
where $E$ generically denotes a heavy charged lepton and $N$ a neutral one, in several 
SM extensions with new leptons in different $\text{SU}(2)_L$ representations. We consider the following additions to the SM particle content:
\begin{itemize}
\item[(1)] A Majorana lepton triplet $\Sigma$, containing a charged lepton $E$ and a Majorana neutrino $N$. Generically, three such triplets appear in minimal seesaw III realisations, but as in previous studies \cite{delAguila:2008cj} we restrict our calculations to the lightest one.
\item[(2)] A Dirac lepton triplet $\Sigma$. This is an alternative to the minimal seesaw III in which two (quasi-)degenerate Majorana triplets $\Sigma_1$, $\Sigma_2$ with opposite CP parities form a
(quasi-)Dirac triplet $\Sigma$~\cite{delAguila:2008hw}, in analogy with the sometimes called ``inverse'' type-I seesaw with heavy Dirac neutrinos~\cite{Mohapatra:1986ks,Buchmuller:1991tu,Datta:1991mf,Ingelman:1993ve}.
In this case the heavy states are two charged leptons $E_1^-$, $E_2^+$ and a Dirac neutrino $N$. 
\item[(3)] A lepton isodoublet $(N\,E)_{L,R}$, in which the heavy neutrino $N$ is likely to have
(quasi-)Dirac character. A lepton isodoublet cannot generate the operator in Eq.~(\ref{ec:O5}) but this does not exclude the possibility that they exist, independently of the neutrino mass generation mechanism. 
\item[(4)] A charged lepton isosinglet $E$, which can also exist independently of the neutrino mass generation.
\item[(5)] A Majorana singlet $N$ and an extra gauge boson $Z'$. For definiteness we work with an $\Esix$ model with a leptophobic $\Zlp$ boson and heavy Majorana neutrinos \cite{delAguila:2007ua}, assuming that only one of them, with mass $m_N < M_{\Zlp}/2$,
can be produced in $\Zlp$ decays. But the results obtained here are general because the relative rate of the multi-lepton signals produced in $pp \to Z' \to NN$ is only determined by the heavy Majorana neutrino decay channels, and the particular type of $Z'$ boson considered only affects the total production cross section. Therefore, the results shown here can be applied to other models~\cite{Huitu:2008gf,Blanchet:2009bu} with a trivial rescaling.
\item[(6)] A Dirac singlet $N$, formed by two Majorana singlets, and an extra $Z'$ boson. We work with the same model in Ref.~\cite{delAguila:2007ua} but assuming that two Majorana neutrinos form a (quasi-)Dirac neutrino.
\end{itemize}
In all the models enumerated above, one or more heavy lepton pairs $E^+ E^-$, $E^\pm N$, $N N$ can be produced in hadron collisions. Moreover, in the case of lepton doublets and triplets the two states $E$, $N$ are almost degenerate in mass. But because the production processes and especially the heavy lepton decay channels are different in each model, the final state signatures are characteristic and the models can be distinguished already at LHC, without the need of precise measurements at a future collider like ILC. The model discrimination relies on the simultaneous analysis of different final states with two, three and four leptons, and the reconstruction of peaks in invariant mass distributions. We will find that the trilepton signal $\ell^\pm \ell^\pm \ell^\mp$, with $\ell=e,\mu$, is common to all models
introducing a heavy neutrino,\footnote{It can also appear in the case of the $E$ singlet but with a very small branching ratio.}
and has a small SM background. Therefore, it constitutes the golden channel for heavy neutrino searches at LHC, in particular for the search of seesaw messengers. 
Moreover, the study of this channel can also reveal if a heavy neutrino eventually observed belongs to a triplet or not. Its Dirac or Majorana nature can be determined in the like-sign dilepton final state: the observation of a $\ell^\pm \ell^\pm$ signal without missing energy clearly indicates its Majorana character whereas the absence of such signal points towards one of the other models.
Like-sign dileptons but with large missing energy can distinguish Dirac triplets, which give a large and observable signal, from Dirac doublets and singlets which give much smaller ones. 
Finally, the model identification is completed with the analysis of four lepton signals and the search of a resonance in the total invariant mass distribution.

We must point out that the models considered in this paper do not exhaust all possibilities for the addition of heavy leptons and/or new interactions to the SM, but rather constitute the cases in which model discrimination seems hardest because the heavy leptons are nearly degenerate, they always decay into a light lepton plus a $W$, $Z$ or $H$ boson and their charge cannot be measured in hadronic decays. Two possibilities not covered here, in which model discrimination is easier, are:
\begin{itemize}
\item A fourth SM generation of chiral leptons. In this case, the mass splitting between the mass eigenstates is expected to be large to be consistent with precise electroweak data~\cite{Kribs:2007nz,Antipin:2009ks,Frandsen:2009fs}, unlike in the cases examined here. Apart from this distinctive characteristic, the leading decay of the charged lepton would be $E \to W^* N$, so that $E^+ E^-$ production would produce events in which the mass resonances have more jets than in the models discussed here. Moreover, anomaly cancellation requires the presence of new quarks with masses of few hundreds of GeV (or new fermions with the same quantum numbers, giving rise to observable resonances \cite{Frandsen:2009fs}), which would be produced in pairs and observed already with a relatively small luminosity~\cite{Ozcan:2008zz,Burdman:2008qh}.
\item A Majorana neutrino coupling to a new $W'$ boson, as for example those appearing in left-right models. In this case the heavy neutrino can be produced in association with a light lepton,
\begin{equation}
pp \to W' \to \ell N \,.
\end{equation}
This process gives signals with up to three charged leptons in some regions of parameter space~\cite{delAguila:2009bb}, but they can be clearly distinguished from heavy lepton pair production by the reconstruction of only one heavy resonance which, together with an energetic light lepton, produces a Jacobian peak at the $W'$ mass.
\end{itemize}
In this work we do not consider scalar triplet production (seesaw II) either, which gives 
signals kinematically very different with sharp peaks in like-sign dilepton invariant mass distributions. It is interesting, however, to note that in seesaw II the trilepton signals are the most important as well~\cite{delAguila:2008cj}. Multi-lepton signals can also appear in pair production of new quarks~\cite{AguilarSaavedra:2009es} but they can be easily distinguished from the models studied here by mass reconstruction and also by the presence of significant single lepton signals with several $b$-tagged jets.

The structure of this paper is as follows. After this introduction, we briefly summarise in section~\ref{sec:2} the relevant Lagrangian terms, the production processes at LHC and the decay channels of the heavy leptons in the models studied. In section~\ref{sec:3} we describe how the signals and backgrounds are simulated. The results for the $\ell^\pm \ell^\pm \ell^\mp$, $\ell^\pm \ell^\pm$ and
$\ell^+ \ell^+ \ell^- \ell^-$ final states are presented in sections \ref{sec:4}--\ref{sec:6}. Section 7 summarises these results and shows explicitly how the six models considered in this work can be distinguished. We draw our conclusions in section~\ref{sec:8}. In appendix~\ref{sec:a} we give the partial widths for heavy lepton decays in the different models studied.

\section{Description of the models}
\label{sec:2}

Here we present the different models studied in turn, collecting the interactions relevant for our analysis and enumerating the production processes at LHC and the allowed decay channels of the heavy leptons. At the end of this section we summarise the main features and compare among the different models.

\subsection{A Majorana lepton triplet}

The relevant Lagrangian for any number of Majorana triplets $\Sigma_i$ has been previously given in
Ref.~\cite{delAguila:2008cj}, and we follow the notation in that work. For a single triplet $\Sigma$, the interactions of the heavy leptons $E$, $N$ with SM leptons $l$, $\nu_l$ is, at first order in the light-heavy mixing $V_{lN}$,
\begin{align}
\mathcal{L}_W & = - g \left( \bar E \gamma^\mu N \, W_\mu^-
   + \bar N \gamma^\mu E \, W_\mu^+ \right) \notag \\
&  - \frac{g}{\sqrt 2} \left( V_{l N} \, \bar l \gamma^\mu  P_L N \; W_\mu^- 
   + V_{l N}^* \, \bar N \gamma^\mu  P_L l \; W_\mu^+ \right) \notag \\
&  - g \left( V_{lN} \, \bar E \gamma^\mu  P_R \nu_l \; W_\mu^- 
    + V_{lN}^* \, \bar \nu_l \gamma^\mu  P_R E \; W_\mu^+ \right) \,, \displaybreak \notag \\ 
\mathcal{L}_Z & = g c_W \, \bar E \gamma^\mu E \, Z_\mu \notag \\
&  + \frac{g}{2c_W} \, \bar \nu_l \gamma^\mu \left(  
V_{l N} P_L - V_{l N}^* P_R \right) N \; Z_\mu \notag \\ 
&  + \frac{g}{\sqrt 2 c_W} \left( V_{lN} \, \bar l \gamma^\mu P_L E
+ V_{lN}^* \, \bar E \gamma^\mu P_L l \right) Z_\mu \,, \notag \\[1mm] 
\mathcal{L}_\gamma & = e \, \bar E \gamma^\mu E \, A_\mu \,, \notag \\
\mathcal{L}_H & =  \frac{g \, m_N}{2 M_W} \,
\bar \nu_l \left( V_{l N} P_R
+ V_{l N}^* P_L \right) N \; H \nonumber \\
& +\frac{g \, m_E}{\sqrt 2 M_W} \left( V_{lN} \, \bar l  P_R E
+ V_{lN}^* \, \bar E  P_L l \right) H \,.
\label{ec:TintM}
\end{align}
Heavy lepton pairs are produced by the gauge $ZEE$, $\gamma EE$ and $WEN$  interactions,
\begin{align}
& q \bar q  \to Z^* / \gamma^* \to E^+ E^- \,, \nonumber \\
& q \bar q' \to W^* \to E^\pm N \,,
\end{align}
(note that there are no $ZNN$ interactions because the triplet has zero hypercharge, and $NN$ pairs are not produced). The heavy leptons $E$, $N$ decay to SM leptons plus a gauge or Higgs boson:
\begin{align}
& E^+ \to \nu W^+ \,,\quad E^+ \to l^+ Z \,,\quad E^+ \to l^+ H \,, \notag \\
& N \to l^- W^+ \,,\quad N \to l^+ W^- \,,\quad N \to \nu Z \,,\quad N \to \nu H \,.
\end{align}
The two heavy states are nearly degenerate in mass, with a small splitting due to radiative corrections, and the decays between the heavy states are very suppressed \cite{Franceschini:2008pz}.

\subsection{A Dirac lepton triplet}

A (quasi-)Dirac lepton triplet is formed by two (quasi-)degenerate Majorana ones with opposite CP parities.
As it was shown in Ref.~\cite{delAguila:2008hw}, the heavy fields can be redefined in such a way that the Lagrangian is written in terms of two charged leptons $E_1^-$, $E_2^+$ (the fermion is positively charged in the second case) and a Dirac neutrino $N$, and lepton number is conserved up to effects of the order of light neutrino masses. At first order in $V_{lN}$ we have
\begin{align}
\mathcal{L}_W & = - g \left( \, \bar E_1^- \gamma^\mu N -\bar N \gamma^\mu E_2^+
\right) W_\mu^-
  -g \left( \bar N \gamma^\mu E_1^- -\bar E_2^+ \gamma^\mu N \right)  W_\mu^+    \notag \\
&  - \frac{g}{\sqrt 2} \left( V_{l N} \, \bar l \gamma^\mu  P_L N \; W_\mu^- 
   + V_{l N}^{*} \, \bar N \gamma^\mu  P_L l \; W_\mu^+ \right) \notag \\
&  + g \left( V_{lN} \, \bar \nu_l \gamma^\mu  P_L E_2^+ \; W_\mu^- 
    + V_{lN}^{*} \, \bar E_2^+ \gamma^\mu  P_L \nu_l \; W_\mu^+ \right) \,, \displaybreak
      \notag \\[1mm]
\mathcal{L}_Z & = g c_W \left( \, \bar E_1^- \gamma^\mu E_1^-
-\bar E_2^+ \gamma^\mu E_2^+ \right) Z_\mu \notag \\
&  + \frac{g}{2c_W} \left( V_{lN} \, \bar \nu_l \gamma^\mu P_L N
+ V_{lN}^{*} \, \bar N \gamma^\mu P_L \nu_l \right) Z_\mu \notag \\
&  + \frac{g}{\sqrt 2 c_W} \left( V_{lN} \, \bar l \gamma^\mu P_L E_1^-
+ V_{lN}^{*} \, \bar E_1^- \gamma^\mu P_L l \right) Z_\mu \,,\notag \\[1mm]
\mathcal{L}_\gamma & = e \left( \, \bar E_1^- \gamma^\mu E_1^-
-\bar E_2^+ \gamma^\mu E_2^+ \right) A_\mu \,, \notag \\[1mm]
\mathcal{L}_H & =  \frac{g \, m_N}{2 M_W} \left( V_{lN} \, \bar \nu_l  P_R N
+ V_{lN}^{*} \, \bar N  P_L \nu_l \right) H \nonumber \\
& +\frac{g \, m_{E_1}}{\sqrt 2 M_W} \left( V_{lN} \, \bar l  P_R E_1^-
+ V_{lN}^{*} \, \bar E_1^-  P_L l \right) H \,.
\label{ec:TintD}
\end{align}
Heavy lepton pairs are produced by the gauge $ZE_i E_i$, $\gamma E_i E_i$ and $WE_iN$  interactions,
\begin{align}
& q \bar q \to Z^* / \gamma^* \to E_i^+ E_i^- \,, \nonumber \\
& q \bar q' \to W^* \to E_i^\pm N \,,
\end{align}
Since there are two charged fermions instead of only one (and a Dirac neutrino is equivalent to two Majorana ones), the total heavy lepton production cross section is twice larger than for a Majorana triplet. The decays have some differences with respect to a Majorana triplet because $E_1^-$ does not couple to light neutrinos, $E_2^+$ does not couple to light charged leptons and $N$ decays conserve lepton number. Thus, the allowed ones are
\begin{align}
& E_1^- \to l^- Z \,,\quad E_1^- \to l^- H \,,\quad E_2^+ \to \nu W^+ \,, \notag \\
& N \to l^- W^+ \,,\quad N \to \nu Z \,,\quad N \to \nu H \,.
\end{align}

\subsection{A lepton isodoublet}

The Lagrangian for a lepton isodoublet can be found in Ref.~\cite{delAguila:1989rq}. With our notation, the terms involved in heavy lepton production and decay are, at first order in $V_{lN}$,
\begin{align}
\mathcal{L}_W & = - \frac{g}{\sqrt 2} \left( \, \bar E \gamma^\mu N \; W_\mu^-
  + \bar N \gamma^\mu E \; W_\mu^+ \right)  \notag \\
&  - \frac{g}{\sqrt 2} \left( V_{l N} \, \bar l \gamma^\mu  P_R N \; W_\mu^- 
   + V_{l N}^{*} \, \bar N \gamma^\mu  P_R l \; W_\mu^+ \right) \notag \\[1mm]
\mathcal{L}_Z & = -\frac{g}{2 c_W} \left( [-1+2 s_W^2] \bar E \gamma^\mu E
+ \bar N \gamma^\mu N \right) Z_\mu \notag \\
&  + \frac{g}{2 c_W} \left( V_{lN} \, \bar l \gamma^\mu P_R E
+ V_{lN}^{*} \, \bar E \gamma^\mu P_R l \right) Z_\mu \,,\notag \\[1mm]
\mathcal{L}_\gamma & = e \, \bar E \gamma^\mu E \; A_\mu \,, \notag \\[1mm]
\mathcal{L}_H & =  \frac{g \, m_E}{2 M_W} \left( V_{lN} \, \bar l  P_L E
+ V_{lN}^{*} \, \bar E  P_R l \right) H \,.
\label{ec:Dint}
\end{align}
Note that the neutrino is likely to be a Dirac fermion because the renormalisable gauge-invariant doublet mass term
\begin{equation}
\mathcal{L}_\text{mass} = - m_D \, \overline{L_{4R}} L_{4L} + \text{H.c.} \,,
\end{equation}
with $L_4 = (N \, E)^T$,
implies a Dirac mass. Additional Majorana masses may appear from dimension-five operators,
\begin{equation}
\mathcal{L}^5_\text{mass} = \frac{C_{44}^L}{\Lambda} \, \overline{L_{4L}^c} \tilde \phi^* \tilde \phi^\dagger L_{4L} + \frac{C_{44}^R}{\Lambda} \, \overline{L_{4R}^c} \tilde \phi^* \tilde \phi^\dagger L_{4R} \,.
\end{equation}
However, if the physics generating these operators is the same as the one yielding the light neutrino mass operator in Eq.~(\ref{ec:O5}), one would expect that the Majorana mass terms,
of order $C_{ij} v^2/\Lambda$, are much smaller than $m_D$. In this case, $N$ is a (quasi-)Dirac fermion. 

The production processes are the same as for a Majorana triplet but now neutral pairs are also produced because they couple to the $Z$ boson,
\begin{align}
& q \bar q  \to Z^* / \gamma^* \to E^+ E^- \,, \nonumber \\
& q \bar q' \to W^* \to E^\pm N \,, \nonumber \\
& q \bar q  \to Z^* \to \bar N N \,.
\end{align}
A further difference is in the size of the couplings, {\em e.g} the coupling to the $W$ boson is reduced by a factor $\sqrt 2$, and so the $E^\pm N$ cross section is a factor of two smaller. In this model the heavy lepton decays are also different, being allowed only
\begin{align}
& E^+ \to l^+ Z \,,\quad E^+ \to l^+ H \,, \notag \\
& N \to l^- W^+ \,.
\end{align}

\subsection{A charged singlet}

The Lagrangian for a charged lepton isosinglet can also be found in Ref.~\cite{delAguila:1989rq}. In our notation, and at first order in the light-heavy mixing $V_{E\nu_l}$, the relevant terms are
\begin{align}
\mathcal{L}_W & = - \frac{g}{\sqrt 2} \left( V_{E\nu_l} \, \bar E \gamma^\mu  P_L \nu_l \; W_\mu^- 
    + V_{E\nu_l}^* \, \bar \nu_l \gamma^\mu  P_L E \; W_\mu^+ \right) \,, \notag \\
\mathcal{L}_Z & = - \frac{g s_W^2}{c_W} \, \bar E \gamma^\mu E \, Z_\mu \notag \\
&  + \frac{g}{2 c_W} \left( V_{E\nu_l} \, \bar E \gamma^\mu P_L \ell
+ V_{E\nu_l}^* \, \bar l \gamma^\mu P_L E \right) Z_\mu \,, \notag \\[1mm]
\mathcal{L}_\gamma & = e \, \bar E \gamma^\mu E \; A_\mu \,, \notag \\
\mathcal{L}_H & =  -\frac{g \, m_E}{2 M_W} \left( V_{E\nu_l} \, \bar E  P_L l
+ V_{E\nu_l}^* \, \bar l  P_R E \right) H \,
\label{ec:Eint}
\end{align}
In this model only charged lepton pairs are produced,
\begin{align}
& q \bar q \to Z^* / \gamma^* \to E^+ E^- \,,
\end{align}
which later decay in the three possible modes
\begin{align}
& E^+ \to \nu W^+ \,,\quad E^+ \to l^+ Z \quad E^+ \to l^+ H \,.
\end{align}

\subsection{A Majorana neutrino and a $Z'$ boson}

A heavy Majorana neutrino which is a singlet under $\text{SU}(3) \times \text{SU}(2)_L \times \text{U}(1)_Y$ interacts with SM fields via a small mixing $V_{lN}$ with SM fermions (for a detailed derivation of the Lagrangian see for example Ref.~\cite{delAguila:2005pf}). Its interactions are therefore suppressed, being at least of order
$V_{lN}$ ($ZNN$ interactions are of order $V_{lN}^2$). However, the interactions with an extra $Z'$ boson are not suppressed if $N$ is not a singlet under the $Z'$ gauge group $\text{U}(1)'$, and are determined by the heavy neutrino charge $Q$ under this extra $\text{U}(1)'$. At first order in $V_{lN}$, the relevant Lagrangian is
\begin{eqnarray}
\mathcal{L}_W & = & - \frac{g}{\sqrt 2}  \left( V_{l N} \, \bar l \gamma^\mu P_L N \; W_\mu^- 
  + V_{l N}^*  \, \bar N \gamma^\mu  P_L l \; W_\mu^+ \right) \,, \nonumber \\
\mathcal{L}_Z & = & - \frac{g}{2 c_W} \, \bar \nu_l \gamma^\mu \left(  
V_{l N} P_L - V_{l N}^* P_R \right) N \; Z_\mu \,, \nonumber \\[1mm]
\mathcal{L}_H & = & - \frac{g \, m_N}{2 M_W} \, \bar \nu_l \left( V_{l N} P_R
+ V_{l N}^* P_L \right) N \; H \,,\notag \\[1mm]
\mathcal{L}_{Z'} & = & - g' \frac{Q}{2}\, \bar N \gamma^\mu \gamma_5 N \; Z'_\mu \,.
\label{ec:NintM}
\end{eqnarray}
In the model considered here~\cite{delAguila:2007ua} we have $Q=-3$, and we have generically denoted the $\text{U}(1)'$ coupling as $g'$. We assume that it equals the SM coupling $g_Y$ of $\text{U}(1)_Y$, but this may change by renormalisation group evolution effects. Heavy neutrino pairs can be produced with the exchange of an $s$-channel $Z'$ boson,
\begin{align}
& q \bar q \to Z' \to N N \,,
\end{align}
and the heavy neutrinos decay giving a $W$, $Z$ or $H$ boson plus a light lepton,
\begin{align}
& N \to l^- W^+ \,,\quad N \to l^+ W^- \,,\quad N \to \nu Z \,,\quad N \to \nu H \,.
\end{align}

\subsection{A Dirac neutrino and a $Z'$  boson}

Finally, in the case of a (quasi-)Dirac neutrino composed by two (quasi-)degenerate Majorana neutrinos with opposite CP parities and equal charges $Q$ under $\text{U}(1)'$, the relevant Lagrangian is
\begin{eqnarray}
\mathcal{L}_W & = & - \frac{g}{\sqrt 2}  \left( V_{l N} \, \bar l \gamma^\mu P_L N \; W_\mu^- 
  + V_{l N}^*  \, \bar N \gamma^\mu  P_L l \; W_\mu^+ \right) \,, \nonumber \\
\mathcal{L}_Z & = & - \frac{g}{2 c_W} \left( V_{l N} \, \bar \nu_l \gamma^\mu  P_L N
   + V_{l N}^* \, \bar N \gamma^\mu P_L \nu_l \right) Z_\mu \,, \nonumber \\[1mm]
\mathcal{L}_H & = & - \frac{g \, m_N}{2 M_W} \left( V_{l N} \, \bar \nu_l  P_R N
+ V_{l N}^* \, \bar N  P_L \nu_l \right) H \,, \notag \\[1mm]
\mathcal{L}_{Z'} & = & - g' Q \, \bar N \gamma^\mu \gamma_5 N \; Z'_\mu
\label{ec:NintD}
\end{eqnarray}
Heavy Dirac neutrino pairs can be produced in the same process
\begin{align}
& q \bar q \to Z' \to \bar N N \,,
\end{align}
and the cross section is twice larger than for a Majorana $N$ because the symmetry factor for identical particles is not present in this case (the $Z' \to \bar N N$ width is also a factor of two larger).\footnote{Note that the Feynman rule for Majorana fermions contains an extra factor of two to account for the two possible Wick contractions, so that the $Z' NN$ vertex is $-i g' Q \gamma^\mu \gamma_5$ as in the Dirac case.} Heavy neutrino decays are the same except for the absence of the LNV one,
\begin{align}
& N \to l^- W^+  \,,\quad N \to \nu Z \,,\quad N \to \nu H \,.
\end{align}

\subsection{Summary}

\begin{figure}[t]
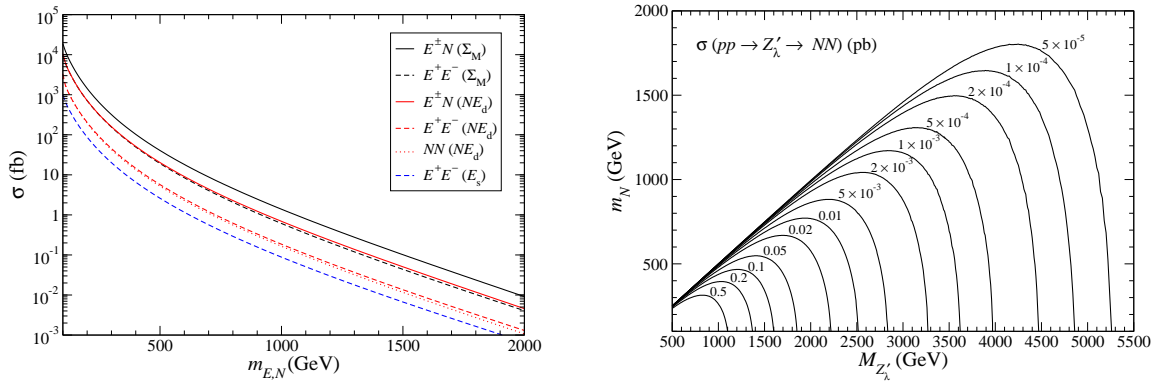

\begin{center}
\begin{tabular}{ccc}
\epsfig{file=Figs/cross-fermion.eps,height=5cm,clip=} & \quad &
\epsfig{file=Figs/contour-Zp.eps,height=5cm,clip=} \\[2mm]
\end{tabular}
\caption{Left: Cross section for heavy lepton pair production in different models. For the Dirac triplet ($\SD$) the cross sections are twice larger than for the Majorana one ($\SM$). Right: cross section for heavy Majorana neutrino pair production from $\Zlp$ decays. For Dirac neutrinos the cross section is two times larger.}
\label{fig:cross}
\end{center}
\end{figure}

In each of the preceding subsections we have indicated the heavy lepton pair production processes present in each model. For better comparison of the different models, we summarise in Table~\ref{tab:g} the coupling constants appearing in the production vertices involved, coupling two heavy leptons to a gauge boson. From now on we will use the labels $\SM$, $\SD$ to refer to the models with Majorana and Dirac triplets, respectively, $\ENd$ and $\Es$ for the doublet $(N\,E)_{L,R}$
and singlet $E_{L,R}$, and $\NM$, $\ND$ for the SM extensions with a leptophobic $\Zlp$ boson and a Majorana or Dirac neutrino. The production cross sections are presented in Fig.~\ref{fig:cross} as a function of the heavy lepton mass (as well as the $Z'$ mass in the corresponding models).
\begin{table}[t]
\begin{center}
\begin{tabular}{cccccc}
                    & $WEN$       & $ZEE$   & $\gamma EE$ & $ZNN$ & $Z'NN$ \\
$\SM$         & $g$         & $g c_W$ & $e$         & 0     & -- \\
$\SD$, $E_1^-$ & $g$         & $g c_W$ & $e$         & 0     & -- \\
$\SD$, $E_2^+$ & $g$         & $g c_W$ & $e$         & 0     & -- \\
$\ENd$        & $g/\sqrt 2$ & $g/c_W (s_W^2-1/2)$ & $e$ & $g/(2 c_W)$ & -- \\
$\Es$         & -- & $gs_W^2/c_W$     & $e$         & -- & --\\
$\NM$         & --          & --      & --          & $\sim 0$ & $g'Q/2$ \\
$\ND$         & --          & --      & --          & $\sim 0$ & $g'Q$ \\
\end{tabular}
\end{center}
\caption{Coupling constants of the gauge vertices involved in heavy lepton pair production,
for the different models considered (the notation is given in the text). The labels $E_1^-$, $E_2^+$ refer to the two charged leptons in the model with a Dirac triplet. A dash indicates that some of the particles in a vertex does not exist in a given model. A zero indicates that the particles exist but they do not couple.}
\label{tab:g}
\end{table}
\begin{table}[t]
\begin{center}
\begin{tabular}{ccccc}
                & $E^+ \to \nu W^+$ & $E^+ \to l^+ Z$ & $E^+ \to l^+ H$ \\
\hline
$\SM$           & 0.5               & 0.25            & 0.25 \\
$\SD$, $E_1^-$   & 0                 & 0.5             & 0.5 \\
$\SD$, $E_2^+$   & 1                 & 0               & 0 \\
$\ENd$          & 0                 & 0.5             & 0.5 \\
$\Es$           & 0.5               & 0.25            & 0.25 \\[2mm]
       & $N \to \ell^- W^+$ & $N \to \ell^+ W^-$ & $N \to \nu Z$ & $N \to \nu H$ \\
\hline
$\SM$  & 0.25               & 0.25               & 0.25          & 0.25 \\
$\SD$  & 0.5                & 0                  & 0.25          & 0.25 \\
$\ENd$ & 1                  & 0                  & 0             & 0 \\
$\NM$  & 0.25               & 0.25               & 0.25          & 0.25 \\
$\ND$  & 0.5                & 0                  & 0.25          & 0.25
\end{tabular}
\end{center}
\caption{Decay branching ratios of the heavy leptons in the limit $m_{E,N} \gg M_W,M_Z,M_H$, for the different models considered (the notation is given in the text). The labels $E_1^-$, $E_2^+$ refer to the two charged leptons in the model with a Dirac triplet.}
\label{tab:BR}
\end{table}
The decay of the heavy leptons takes place in the channels indicated, with partial widths collected in appendix~\ref{sec:a} for reference. Nevertheless, the important quantities for LHC phenomenology are the relative branching ratios. Summing over light leptons $l=e,\mu,\tau$, the branching ratios into $W$, $Z$ and $H$ bosons are independent of the mixing. Their values for $m_{E,N} \gg M_W,M_Z,M_H$, are
collected in Table~\ref{tab:BR}. This table illustrates the important differences among the models considered, which make the discrimination based on multi-lepton signals quite effective. Note that for masses $m_{E,N} = 300$ GeV as considered in the following sections, the branching ratios are already close to the values presented in this table.

\section{Multi-lepton signal generation}
\label{sec:3}

The analysis pursued in this work, aiming to discriminate among several models all giving various multi-lepton signals in different decay channels, is somewhat demanding from the point of view of the simulation. It requires to generate {\em all} the signal contributions because many different heavy lepton decay channels, with the subsequent $W/Z/H$ decay, can lead to the same charged lepton multiplicities. The complete signal generation has been done with the program {\tria}~\cite{delAguila:2008cj} extended to include the models with an isodoublet $(N\,E)_{L,R}$, an isosinglet $E_{L,R}$ and $Z'$ production with decay to a Majorana or Dirac neutrino. All the signal processes enumerated in the previous section, with all the possible decays of the heavy leptons $E$, $N$ and the $W/Z$ bosons, are included. The Higgs boson decay, which does not carry any spin information, is left to the parton shower Monte Carlo. Signals have been generated with statistics of 300 fb$^{-1}$ and rescaled to a reference luminosity of 30 fb$^{-1}$, in order to reduce statistical fluctuations.
The SM background, consisting of the processes in Table~\ref{tab:bkg},
is generated using {\tt Alpgen}~\cite{Mangano:2002ea} with a Monte Carlo statistics of 30 fb$^{-1}$. 
Additional SM processes which were previously shown to be negligible after selection cuts~\cite{delAguila:2008cj} are ignored in this work. 
Signals and backgrounds are passed through the parton shower Monte Carlo {\tt Pythia} 6.4\cite{Sjostrand:2006za} to add initial and final state radiation (ISR, FSR) and pile-up, and perform hadronisation.
For the backgrounds we use the MLM prescription~\cite{mlm} for the matching to avoid double counting between the matrix-level generator and the parton shower Monte Carlo. We use the fast simulation {\tt AcerDET}~\cite{RichterWas:2002ch} which is a generic LHC detector simulation, neither of ATLAS nor of CMS, finding good agreement with our previous results~\cite{delAguila:2008cj}.

\begin{table}[htb]
\begin{center}
\begin{small}
\begin{tabular}{llc}
Process & Decay & Events \\
\hline
$t \bar t nj$, $n=0,\dots,5$   & semileptonic           & 6.1 M \\
$t \bar t nj$, $n=0,\dots,5$   & dileptonic             & 1.5 M \\
$tW$                           & all                    & 1.6 M   \\
$W t \bar t nj$, $n=0,\dots,4$ & $W \to l \nu$          & 5.1 K \\
$Z b \bar b nj$, $n=0,\dots,4$ & $Z \to l^+ l^-$        & 200 K \\
$Z t \bar t nj$, $n=0,\dots,4$ & $Z \to l^+ l^-$        & 1.87 K \\
$WWnj$, $n=0,\dots,3$          & $W \to l \nu$          & 290 K \\
$WZnj$, $n=0,\dots,3$          & $W \to l \nu$, $Z \to l^+ l^-$
                                                        & 37.7 K \\
$ZZnj$,  $n=0,\dots,3$         & $Z \to l^+ l^-$        & 3.74 K \\
$WWWnj$, $n=0,\dots,3$         & $2W \to l \nu$         & 1.47 K \\
\end{tabular}
\end{small}
\caption{Background processes considered in the simulations, with $nj$ standing for $n$ additional jets at the partonic level. The second column indicates the decay modes included (where $l=e,\mu,\tau$). The last column corresponds to the number of events after matching for a luminosity of 30 fb$^{-1}$, with K and M standing for $10^3$ and $10^6$ events, respectively.}
\label{tab:bkg}
\end{center}
\end{table}

We remark that the use of at least a fast simulation of the detector is essential for this and other similar studies, because some of the most important SM backgrounds have charged leptons resulting from $b$ quark decays, and these cannot be estimated in a parton-level analysis. In fact, the most recent analyses for supersymmetry searches performed with a full detector simulation~\cite{Aad:2009wy} confirm the well-known fact that $t \bar t nj$ is probably the largest (and most dangerous) SM source of like-sign dileptons. From the point of view of the signal, the use of a fast detector simulation is also necessary because some of the contributions to the multi-leptonic final states studied arise when more charged leptons are produced but missed by the detector. 

In the following sections we will take heavy lepton masses $m_{E,N} = 300$ GeV for our simulations. 
Production cross sections are independent of the mixing and, except for unnaturally small values, the heavy leptons will decay well inside the detector. Note that the heavy lepton widths are larger than the $b$ quark width for $V \gtrsim 10^{-7}$, and present indirect constraints are of order $V \lesssim 10^{-1}$~\cite{delAguila:2008pw}.
In definite seesaw models, heavy lepton mixing is also related to light neutrino masses. However, we do not assume any particular model-dependent relation between light neutrino masses and heavy lepton mixing.
Instead, we assume that heavy leptons only couple to the first generation, bearing in mind that
for an arbitrary mixing with the first two generations the results obtained would be approximately the same (the signals are equivalent and at high transverse momenta the SM backgrounds involving electrons and muons have roughly the same size). On the other hand, if $E$, $N$ only mix with the tau the signals would be very difficult to observe~\cite{delAguila:2008cj}.
For the $\Zlp$ boson we will conservatively take a mass of 650 GeV, because this is approximately the location of the maximum in the heavy lepton pair invariant mass distribution for $m_{E,N} = 300$ GeV when they are produced by off-shell $W/Z$ bosons. Hence, for such mass the identification of an $s$-channel resonance would be more difficult. This $\Zlp$ mass is not excluded by present Tevatron searches for $t \bar t$ resonances. For the $\Zlp$ model considered, the cross section into $t \bar t$ final states (assuming $g'=g_Y$) is about 0.2 pb, well below the 95\% confidence level limit for this mass, which is around 0.6 pb~\cite{ttres}. The Higgs boson mass is taken as $M_H = 115$ GeV, in which case it mainly decays into two jets,
$H \to b \bar b,c \bar c,gg$ and seldom produces leptons, only when $H \to \tau^+ \tau^-$ with $\tau$ leptonic decay. For a heavier Higgs decaying into $W^+ W^-$, $ZZ$, the multi-lepton signals examined  here would still be present but some signals with higher lepton multiplicity,
originating from heavy lepton decays involving a Higgs boson and $H \to W^+ W^-,ZZ$, would also be present and might be of interest. In addition, several SM backgrounds (for example, $W^\pm H \to W^\pm W^+ W^-$ plus jets) would be enhanced. A dedicated analysis is required to examine the precise discovery potential of each channel in such case.

Finally, it is worth commenting here about the statistical prescriptions used to determine a possible discovery. If the background can be precisely known or directly estimated from data, for instance if the signal shows up as a sharp peak, the statistical significance is $\mathcal{S}_0 = S/\sqrt B$, where $S$, $B$ are the number of signal and background events, respectively. For small $B$, this estimator is replaced by the $P$-number using Poisson statistics. The discovery criteria used in this work are: (i) statistical significance larger than $5\sigma$, and (ii) the presence of at least 10 signal events. In most of our results the luminosity required for discovery does not depend on a very precise background normalisation because the background is tiny, so that the signal significance is well above $5 \sigma$ and the limit is determined by having 10 signal events.


\section{Final state $\ell^\pm \ell^\pm \ell^\mp$}
\label{sec:4}

This final state is the most characteristic one of a heavy neutrino, and it can appear both in $E^\pm N$ and $NN$ production, for example in the decay channels
\begin{align}
& E^+ N \to \ell^+ Z \, \ell^\pm W^\mp \,,
  && \quad Z \to q \bar q / \nu \bar \nu, W \to \ell \nu \,, \nonumber \\
& E^+ N \to \ell^+ H \, \ell^\pm W^\mp \,,
  && \quad H \to q \bar q , W \to \ell \nu  \,, \nonumber \\
& N N \to \ell^+ W^- \, \ell^- W^+ \,,
  && \quad W W \to q \bar q \ell \nu \,,
\label{ec:ch3Q1}
\end{align}
irrespectively of the Dirac or Majorana character of $N$. For a lepton triplet, it can also be produced in several other decays, for example
\begin{align}
& E^+ N \to \ell^+ Z \, \nu Z \,,
  && \quad ZZ \to \ell^+ \ell^- \, q \bar q/ \nu \bar \nu \,, \nonumber \\
& E^+ N \to \ell^+ Z \, \nu H / \ell^+ H \, \nu Z \,,
  && \quad Z \to \ell^+ \ell^- , H \to q \bar q \,.
\label{ec:ch3Q1b}
\end{align}
Note that $N \to \nu Z$, $N \to \nu H$ do not take place in the case of a lepton doublet, which can be exploited to distinguish them from triplets. With this aim, 
we perform two complementary analyses for the $\ell^\pm \ell^\pm \ell^\mp$ final state, dividing the sample into two disjoint ones. First, we perform a generic analysis as in Ref.\cite{delAguila:2008cj}, which can detect the presence of a heavy neutrino singlet, doublet or triplet decaying in the channels of Eqs.~(\ref{ec:ch3Q1}). 
In this analysis we reject events with a $Z$ boson candidate, that is, with an opposite-charge lepton pair with an invariant mass consistent with $M_Z$. This sample is labelled as $\ell^\pm \ell^\pm \ell^\mp$ (no $Z$).
In second place, we perform a new specific analysis to search for the decays in Eqs.~(\ref{ec:ch3Q1b}) and determine whether the neutrino belongs to a triplet. In this case we only accept events with a $Z$ boson candidate. This sample is labelled as $\ell^\pm \ell^\pm \ell^\mp$ ($Z$).
The common event pre-selection criteria for the two analyses are: (i) the presence of two like-sign leptons $\ell_1$ and $\ell_2$ with transverse momentum $p_T > 30$ GeV and an additional (and only one) lepton of opposite sign, with $p_T > 10$ GeV; (ii) two hard jets with $p_T > 20$ GeV.
The cut on transverse momenta of the like-sign pair greatly reduces the SM background from $t \bar tnj$ production. The requirement of two jets is used in the kinematical reconstruction, and also reduces the $WZnj$ background. We must point out that
this background, as simulated with {\tt Alpgen}, does not include off-shell photons and uses the narrow width approximation for both bosons. The effect of off-shell photons is important in general~\cite{Sullivan:2008ki} but in this case this contribution is reduced by the high-$p_T$ requirement on charged leptons. More important is the effect of the $Z$ width: the narrow approximation underestimates the background for the `no $Z$' sample and overestimates it in the complementary `$Z$' one. With a comparison of the reconstructed $\ell^+ \ell^-$ distributions for $Z/\gamma^*$ and $Z$ on-shell after detector simulation, we estimate that for the `no $Z$' trilepton final state the background can be at most two times larger than the values given here. This can be compensated, at the cost of some signal efficiency loss, by a wider interval for rejection of events with a $Z$ candidate, and the discovery potential would be very similar. Model discrimination, of course, would not be affected.

\subsection{Final state $\ell^\pm \ell^\pm \ell^\mp$ (no $Z$)}
\label{sec:4.1}

In this analysis we ask for event selection the absence of a $Z$ boson candidate: neither of the two opposite-sign lepton pairs can have an invariant mass closer to $M_Z$ than 10 GeV. This is very useful to remove $WZnj$ production but also eliminates several of the signal channels. We collect in Table~\ref{tab:nsnb-3Q1} the number of signal and background events for each process and model after pre-selection and selection cuts.

\begin{table}[htb]
\begin{center}
\begin{tabular}{cccccccccc}
                      & Pre.  & Sel.  & Peak & \quad &                & Pre. & Sel.  & Peak \\[1mm]
$E^+ E^-$ ($\SM$)     & 58.1  & 26.3  & 5.7   & & $E^+ E^-$ ($\ENd$) & 38.3  & 23.7  & 5.4 \\
$E^\pm N$ ($\SM$)     & 269.2 & 192.2 & 86.3  & & $E^\pm N$ ($\ENd$) & 393.2 & 355.1 & 183.8 \\
$E_1^+ E_1^-$ ($\SD$) & 127.2 & 80.9  & 20.0  & & $N N$ ($\ENd$)     & 164.4 & 155.7 & 87.8 \\
$E_2^+ E_2^-$ ($\SD$) & 0.0   & 0.0   & 0.0   & & $E^+ E^-$ ($\Es$)  & 8.2   & 3.1   & 0.7 \\
$E_1^\pm N$ ($\SD$)   & 502.1 & 370.2 & 181.9 & & $N N$ ($\NM$)      & 311.0 & 252.6 & 143.2 \\
$E_2^\pm N$ ($\SD$)   & 36.1  & 28.1  & 3.3   & & $N N$ ($\ND$)      & 576.2 & 481.9 & 285.5 \\
\hline
$t \bar t nj$         & 236   & 156   & 0     & & $WZnj$             & 1540  & 38    & 2 \\
$W t \bar t nj$       & 54    & 47    & 6     & & $ZZnj$             & 86    & 5     & 0 \\
$Z t \bar t nj$       & 151   & 20    & 3     & & $WWWnj$            & 17    & 12    & 3 \\
\end{tabular}
\end{center}
\caption{Number of events in the $\ell^\pm \ell^\pm \ell^\mp$ (no $Z$) sample for
the signals and main backgrounds with a luminosity of 30 fb$^{-1}$.}
\label{tab:nsnb-3Q1}
\end{table}

The event reconstruction is performed in three steps, following this procedure~\cite{delAguila:2008cj}:
\begin{enumerate}
\item The momentum of the $Z$ or $H$ boson decaying hadronically is reconstructed as the sum of the momenta of the leading and sub-leading jets.
\item One of the heavy charged leptons $L$ ($E$ or $N$, depending on the process) can be reconstructed from this boson and one of the two like-sign leptons, and the heavy neutrino $N$ from the two remaining leptons (with opposite charge) and the missing neutrino momentum. The longitudinal component of the neutrino momentum is neglected for the moment, and the transverse component is taken as the missing energy. There are two possibilities for this pairing, and we choose the one giving closest invariant masses for the reconstructed $L$ and $N$.
\item The $N$ reconstruction can be refined by including the longitudinal neutrino momentum. We select among the two charged leptons the least energetic one $\ell_\text{s}$, and require that its invariant mass with the neutrino is $M_W$,
\begin{equation}
(p_{\ell_\text{s}} + p_\nu)^2 = M_W^2 \,,
\label{ec:pnurec3l}
\end{equation}
taking the transverse components of $p_\nu$ as the missing energy.
This quadratic equation determines the longitudinal neutrino momentum up to a twofold ambiguity, which is resolved selecting the solution with smaller $(p_\nu)_z$. In case that no real solution exists, the transverse neutrino momentum 
used in Eq.~(\ref{ec:pnurec3l}) is decreased until a real solution is found.
\end{enumerate}

\begin{figure}[t]
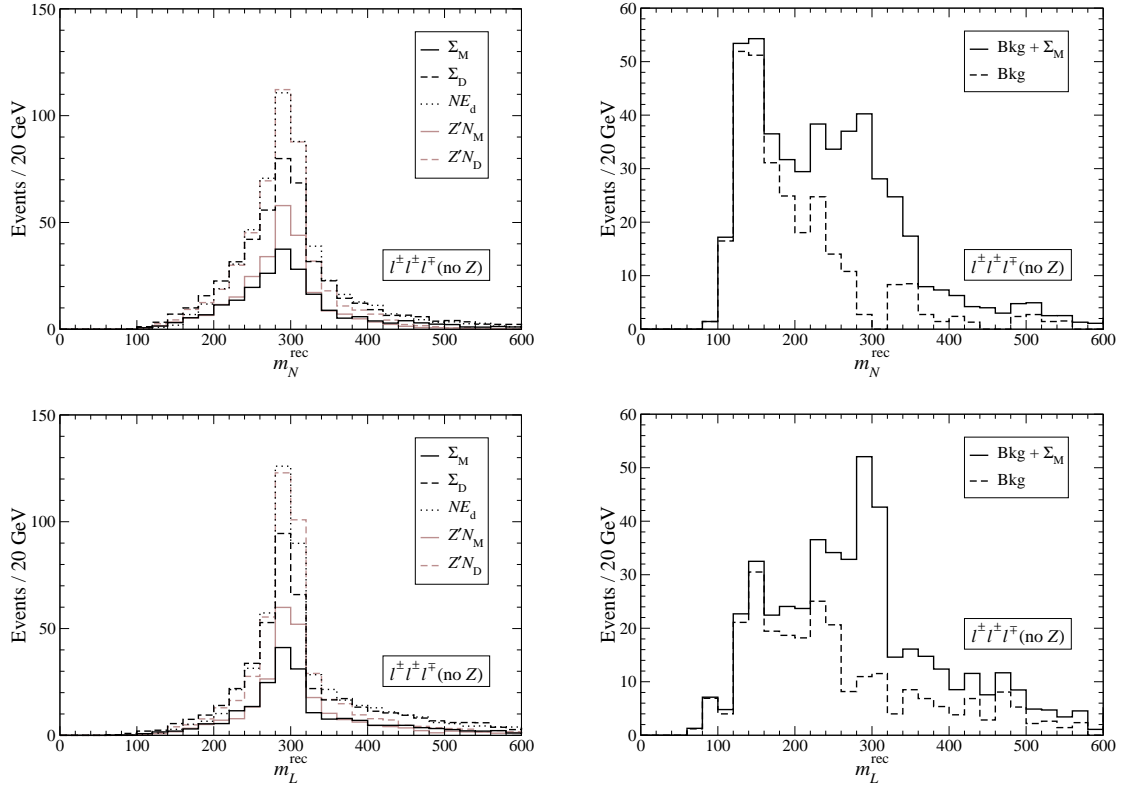

\begin{center}
\begin{tabular}{ccc}
\epsfig{file=Figs/mN-3Q1.eps,height=5cm,clip=} & \quad &
\epsfig{file=Figs/mN-3Q1-BS1.eps,height=5cm,clip=} \\[2mm]
\epsfig{file=Figs/mL-3Q1.eps,height=5cm,clip=} & \quad &
\epsfig{file=Figs/mL-3Q1-BS1.eps,height=5cm,clip=} 
\end{tabular}
\caption{Left: reconstructed heavy lepton masses for the signals
in the $\ell^\pm \ell^\pm \ell^\mp$ (no $Z$) sample. Right: the same for the SM background and the background plus the Majorana triplet signal. The luminosity is 30 fb$^{-1}$.}
\label{fig:mrec-3Q1}
\end{center}
\end{figure}

The reconstructed heavy lepton masses after selection criteria are shown in Fig.~\ref{fig:mrec-3Q1} (left) for the five models which give an observable signal. We point out that one of the resonances, labelled as $N$, can be identified as being a neutral lepton, because the peak appears in the invariant mass distribution of two opposite-charge leptons plus missing energy. The identity of the other resonance, labelled as $L$, cannot be established because the charge of the hadronic jets cannot be measured. In fact, depending on the process and the model, it can be a charged or neutral heavy lepton (see Table~\ref{tab:nsnb-3Q1}). In Fig.~\ref{fig:mrec-3Q1} (right) we show the same distributions for the SM background and the background plus the Majorana triplet signal, which is the smallest one. We define the peaks as the intervals
\begin{align}
240 < m_L^\text{rec} < 360~\text{GeV} \,, \notag \\
240 < m_N^\text{rec} < 360~\text{GeV} \,,
\end{align}
and show in Table~\ref{tab:nsnb-3Q1} the number of signal and background events after these cuts.
The statistical significance for 30 fb$^{-1}$ of the relevant signals (neglecting the background uncertainty) is collected in Table \ref{tab:sig-3Q1}, with the luminosity required to achieve $5\sigma$ discovery. Due to the smallness of the background, the discovery luminosity is determined in each case by the requirement of having at least 10 signal events. For such small luminosities the SM background is tiny, around one event or less, and a very precise normalisation is not very important.
Notice also that the signals (and thus the statistical significance) are larger for a lepton doublet than for the triplets, despite the smaller production cross section. This is mainly due to the larger (100\%) branching ratio of the mode $N \to \ell^- W^+$, and also to the presence of $\bar N N$ production.

\begin{table}[ht]
\begin{center}
\begin{tabular}{ccccccc}
       & $\mathcal{S}_0$ & $L$       & \quad &     & $\mathcal{S}_0$ & $L$      \\[1mm]
$\SM$  & 24.6        & 3.3 fb$^{-1}$ & & $\NM$ & 38.3 & 2.1 fb$^{-1}$ \\
$\SD$  & 54.8        & 1.5 fb$^{-1}$ & & $\ND$ & 76.3 & 1.1 fb$^{-1}$ \\
$\ENd$ & 74.0        & 1.1 fb$^{-1}$ & & 
\end{tabular}
\end{center}
\caption{Statistical significance of the relevant signals for 30 fb$^{-1}$, and luminosity $L$ required to have a $5\sigma$ discovery in the $\ell^\pm \ell^\pm \ell^\mp$ (no $Z$) sample.}
\label{tab:sig-3Q1}
\end{table}

\begin{figure}[htb]
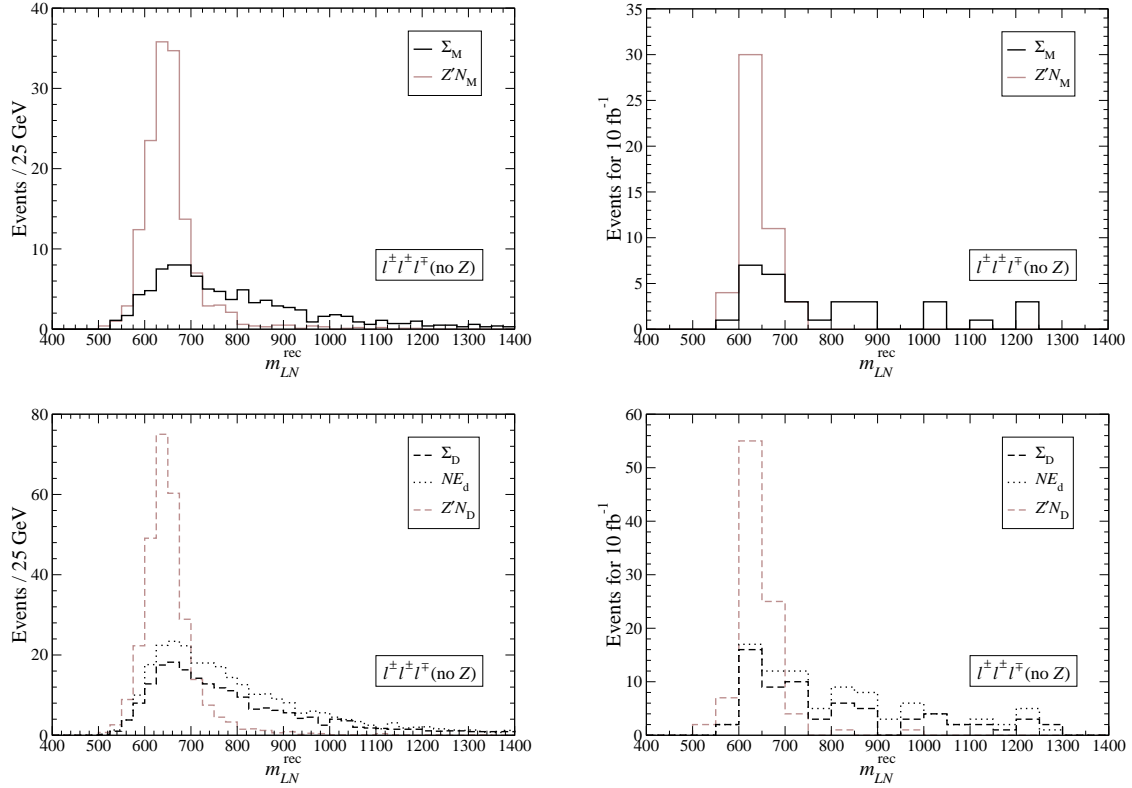

\begin{center}
\begin{tabular}{ccc}
\epsfig{file=Figs/mLN-3Q1-M.eps,height=5cm,clip=} & \quad &
\epsfig{file=Figs/mLN-3Q1-MX.eps,height=5cm,clip=} \\[2mm]
\epsfig{file=Figs/mLN-3Q1-D.eps,height=5cm,clip=} & \quad &
\epsfig{file=Figs/mLN-3Q1-DX.eps,height=5cm,clip=} 
\end{tabular}
\caption{Left: invariant mass distribution of the heavy lepton pair in the $\ell^\pm \ell^\pm \ell^\mp$ (no $Z$) final state, after reconstruction cuts, normalised to 30 fb$^{-1}$. Right: possible experimental result for 10 fb$^{-1}$, where the numbers of signal events are 31 ($\SM$), 68 ($\SD$), 92 ($\ENd$),
48 ($\NM$) and 95 ($\ND$).}
\label{fig:mLN-3Q1}
\end{center}
\end{figure}

With the results shown,
we observe that this final state is very sensitive to the presence of heavy neutrinos in all the models considered (except the one with the charged singlet, which does not have a neutral lepton).\footnote{Note that the sensitivity to Majorana triplets can be improved with a more inclusive analysis without event reconstruction, and the luminosity required for $5\sigma$ discovery can be reduced to 1.7 fb$^{-1}$~\cite{delAguila:2008cj}.}
 We can then ask ourselves whether one could already distinguish some of these models. This is indeed possible, although model discrimination probably requires more luminosity than $5\sigma$ discovery. By examination of the heavy lepton pair invariant mass, one can determine if these leptons are produced by the exchange of an $s$-channel resonance, as it is the case in the models with an extra $\Zlp$ boson. We show in Fig.~\ref
{fig:mLN-3Q1} this kinematical distribution for the signals only, separating for clarity the models in which the heavy neutrino has Majorana (up) and Dirac nature (down), which can be distinguished by other means (see section~\ref{sec:5.1}).
On the right we show possible experimental results for 10 fb$^{-1}$, obtained by making random fluctuations with a Poisson distribution of the bins in the distributions, and normalising to the total expected number of events. The background is much smaller than the signals and has not been included. It is quite clear that the presence of a resonance can be detected or excluded, possibly with a smaller luminosity, but we do not address this issue quantitatively.

\subsection{Final state $\ell^\pm \ell^\pm \ell^\mp$ ($Z$)}
\label{sec:4.2}

The trilepton sample with a $Z$ boson candidate suffers from a large $WZnj$ background. Nevertheless,
several of the decay channels in Eqs.~(\ref{ec:ch3Q1b}) produce a sharp peak in the trilepton invariant mass distribution, corresponding to the heavy charged lepton mass $m_E$. This peak can be observed over the large background. As event selection criteria we require: (i) a $Z$ boson candidate, with two opposite-charge leptons having an invariant mass between $M_Z-10$ GeV and $M_Z + 10$ GeV; (ii) missing energy $\ptmiss > 30$ GeV. The number of signal and background events after event pre-selection and selection  is given in Table~\ref{tab:nsnb-3Q1b}.
\begin{table}[t]
\begin{center}
\begin{tabular}{cccccccccc}
                      & Pre.  & Sel.  & Peak & \quad &                & Pre. & Sel.  & Peak \\[1mm]
$E^+ E^-$ ($\SM$)     & 58.1  & 24.3  & 20.5  & & $E^+ E^-$ ($\ENd$) & 38.3  & 6.0   & 1.8 \\
$E^\pm N$ ($\SM$)     & 269.2 & 66.6  & 26.5  & & $E^\pm N$ ($\ENd$) & 393.2 & 21.2  & 4.1 \\
$E_1^+ E_1^-$ ($\SD$) & 127.2 & 15.7  & 5.3   & & $N N$ ($\ENd$)     & 164.4 & 7.8   & 1.1 \\
$E_2^+ E_2^-$ ($\SD$) & 0.0   & 0.0   & 0.0   & & $E^+ E^-$ ($\Es$)  & 8.2   & 3.9   & 3.4 \\
$E_1^\pm N$ ($\SD$)   & 502.1 & 111.3 & 51.3  & & $N N$ ($\NM$)      & 311.0 & 54.6  & 10.6 \\
$E_2^\pm N$ ($\SD$)   & 36.1  & 7.6   & 0.9   & & $N N$ ($\ND$)      & 576.2 & 90.1  & 13.9 \\
\hline
$t \bar t nj$         & 236   & 66    & 3     & & $WZnj$             & 1540  & 1063   & 65 \\
$Z t \bar t nj$       & 54    & 101   & 5     & & $ZZnj$             & 86    & 21     & 1 \\
\end{tabular}
\end{center}
\caption{Number of events in the $\ell^\pm \ell^\pm \ell^\mp$ $(Z)$ sample for
the signals and main backgrounds with a luminosity of 30 fb$^{-1}$.}
\label{tab:nsnb-3Q1b}
\end{table}
\begin{figure}[t]
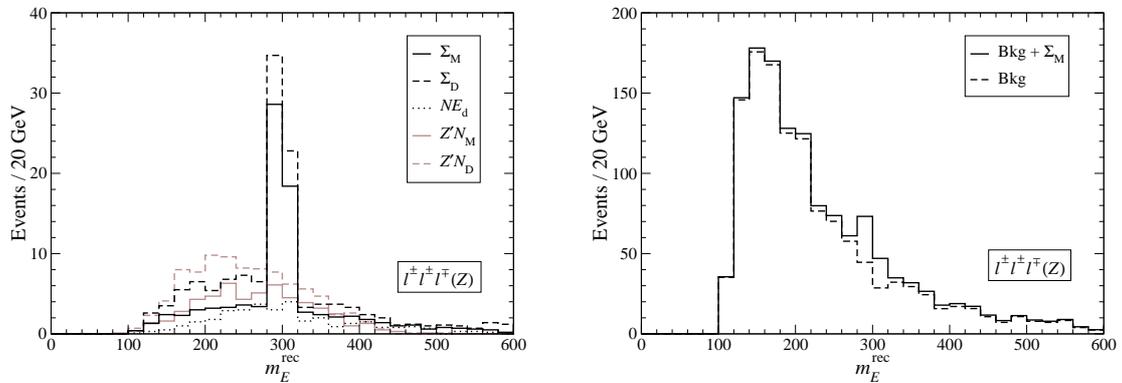

\begin{center}
\begin{tabular}{ccc}
\epsfig{file=Figs/mE-3Q1b.eps,height=5cm,clip=} & \quad &
\epsfig{file=Figs/mE-3Q1b-BS1.eps,height=5cm,clip=}
\end{tabular}
\caption{Left: reconstructed heavy charged lepton mass for the signals
in the $\ell^\pm \ell^\pm \ell^\mp$ ($Z$) sample. Right: the same for the SM background and the background plus the Majorana triplet signal. The luminosity is 30 fb$^{-1}$.}
\label{fig:mrec-3Q1b}
\end{center}
\end{figure}
The heavy $E$ mass is simply reconstructed as the three-lepton invariant mass. Its distribution is shown in Fig.~\ref{fig:mrec-3Q1b} (left) for the five non-negligible signals. 
For the lepton triplet the peaks could be seen over the SM background, as it is shown in the right side of this figure for the Majorana triplet. These results can be understood with the following considerations:
\begin{enumerate}
\item The off-peak contributions in the lepton triplet models result, for example, from the decay channels in Eqs.~(\ref{ec:ch3Q1b}) with $N \to \nu Z \to \nu \ell^+ \ell^-$. In this case the resonance would be seen in the invariant mass of two jets plus the other lepton, all of them being $E$ decay products. However, the corresponding peak is very broad due to the worse jet energy resolution, and difficult to see over the large $WZnj$ background.
\item As expected, in processes where a charged lepton is not present, as in $\Zlp \to NN$, the distributions do not display a peak.
\item For a lepton doublet the decay channel $E N \to \ell Z \, \ell' W$, with $\ell'$ missed by the detector and $Z \to \ell^+ \ell^-$, $W \to q \bar q$, would also produce a trilepton signal with a peak at $m_E$. However, this process is removed by the missing energy requirement, and the rest of the decay channels do not have any resonance in the trilepton invariant mass, as it can be seen in Fig.~\ref{fig:mrec-3Q1b}. This feature clearly distinguishes a lepton doublet from a (Majorana or Dirac) triplet.
\end{enumerate}
Defining the peak region as
\begin{align}
280 < m_E^\text{rec} < 320~\text{GeV} \,,
\end{align}
and performing a kinematical cut on this reconstructed mass, we obtain for each process the number of events listed in Table~\ref{tab:nsnb-3Q1b}. The statistical significance of the signals and the luminosity required for $5\sigma$ discovery are presented in Table~\ref{tab:sig-3Q1b}.
\begin{table}[t]
\begin{center}
\begin{tabular}{ccccccc}
       & $\mathcal{S}_0$ & $L$       & \quad &     & $\mathcal{S}_0$ & $L$      \\[1mm]
$\SM$  & 5.5        & 25 fb$^{-1}$   & & $\SD$  & 6.7        & 16.6 fb$^{-1}$
\end{tabular}
\end{center}
\caption{Statistical significance for 30 fb$^{-1}$ of the relevant signals, and luminosity $L$ required to have a $5\sigma$ discovery in the $\ell^\pm \ell^\pm \ell^\mp$ ($Z$) sample.}
\label{tab:sig-3Q1b}
\end{table}
We assume that the background can be accurately normalised with off-peak measurements of the trilepton invariant mass distribution, as it is apparent in Fig.~\ref{fig:mrec-3Q1b} (right).
It is clear that this sample improves little the global statistical significance of the trilepton channel.\footnote{The sensitivity shown here may be significantly improved by further reducing the background if we ask for the presence of two hard jets and set a more stringent cut on $m_E^\text{rec}$. However, in a real experiment it may be difficult to spot the presence of a peak with small statistics due to the background fluctuations, and to be conservative we have not performed these optimisations.}
But the observation of a positive signal here, with about half the size of the signal in the sample without $Z$ candidates, is a proof of the triplet nature of the heavy neutrino. And it also proves the presence of a charged heavy lepton, although this can be established more easily in the four lepton final state, as it will be shown in section~\ref{sec:6}.


\section{Final state $\ell^\pm \ell^\pm$}
\label{sec:5}

The like-sign dilepton final state without significant missing energy is characteristic of a heavy Majorana neutrino whose decays violate lepton number. It can be produced in decays such as
\begin{align}
& E^+ N \to \ell^+ Z \, \ell^+ W^- \,,
  && \quad Z \to q \bar q, W \to q \bar q' \,, \nonumber \\
& E^+ N \to \ell^+ H \, \ell^+ W^- \,,
  && \quad H \to q \bar q , W \to q \bar q' \,, \nonumber \\
& N N \to \ell^\pm W^\mp \, \ell^\pm W^\mp \,,
  && \quad WW \to q \bar q' q \bar q'  \,.
\label{ec:ch2Q2}
\end{align}
Large like-sign dilepton signals can also arise from lepton number conserving (LNC) processes, most notoriously in
\begin{align}
& E_2^+ \bar N \to \nu W^+ \, \ell^+ W^- \,,
  && \quad W^+ \to \ell^+ \nu , W^- \to q \bar q' \,
\label{ec:ch2Q2b}
\end{align}
in the Dirac triplet model, being lepton number balanced by final state neutrinos (remember that $E_2^+$ is a fermion, not an anti-fermion). They can also appear when more charged leptons are produced but are missed by the detector, for example
\begin{align}
& E^+ N \to \ell^+ Z \, \ell^- W^+ \,,
  && \quad Z \to q \bar q, W \to \ell^+ \nu \,, \nonumber \\
& \bar N N \to \ell^+ W^- \, \ell^- W^+ \,,
  && \quad W^- \to q \bar q' , W^+ \to \ell^+ \nu \,,
\end{align}
with $\ell^-$ missed by the detector. The presence of such signals should not constitute a surprise, since several SM processes, for instance $W^\pm W^\pm nj$ and $WZnj$ production, give like-sign dileptons in this way. 
Hence, as in the trilepton channel we will perform two different analyses with disjoint event samples. In the first one we will, as in Ref.~\cite{delAguila:2008cj}, impose selection criteria to isolate the truly LNV processes characteristic of a Majorana fermion. In particular, we will require the absence of significant missing energy. In second place we will perform a new analysis for events with large missing energy, aiming to
isolate the large signal from the decay channel in Eq.~(\ref{ec:ch2Q2b}). The common pre-selection will be to ask the presence of two like-sign leptons $\ell_1$ and $\ell_2$ with transverse momentum $p_T > 30$ GeV.

\subsection{Final state $\ell^\pm \ell^\pm$ (no $\ptmiss$)}
\label{sec:5.1}

\begin{figure}[htb]
\begin{center}
\begin{tabular}{ccc}
\epsfig{file=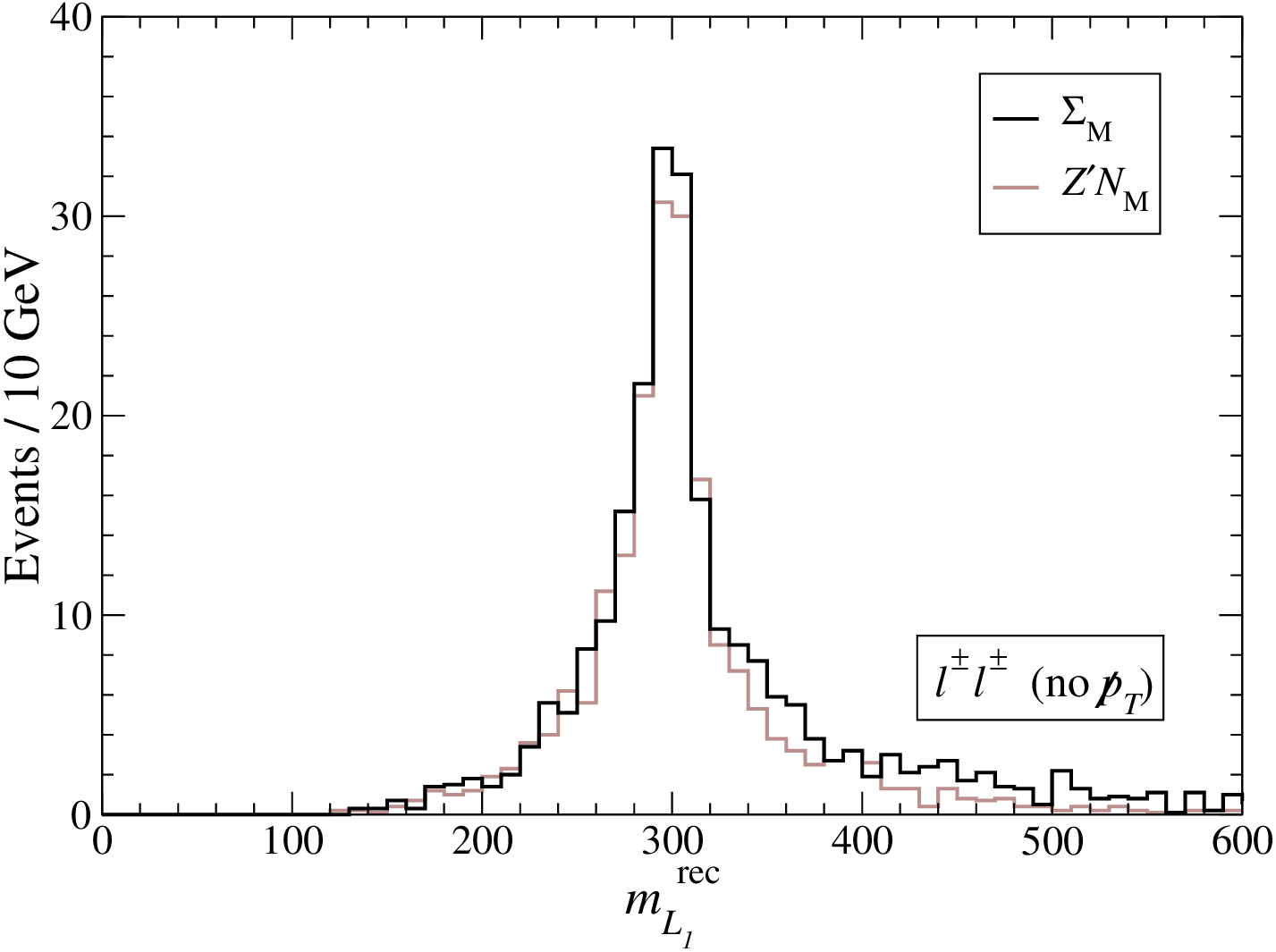,height=5cm,clip=} & \quad &
\epsfig{file=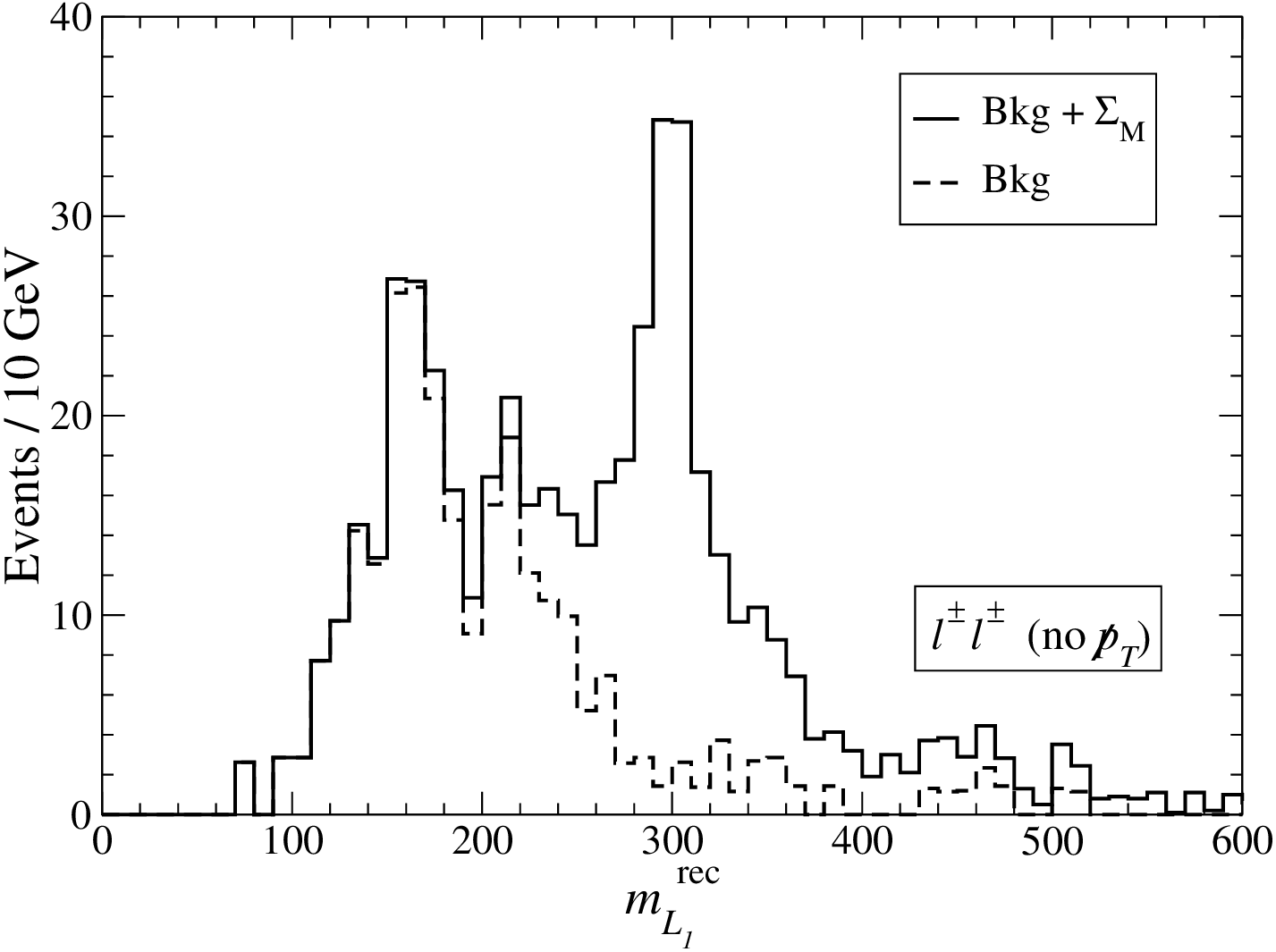,height=5cm,clip=} \\[2mm]
\epsfig{file=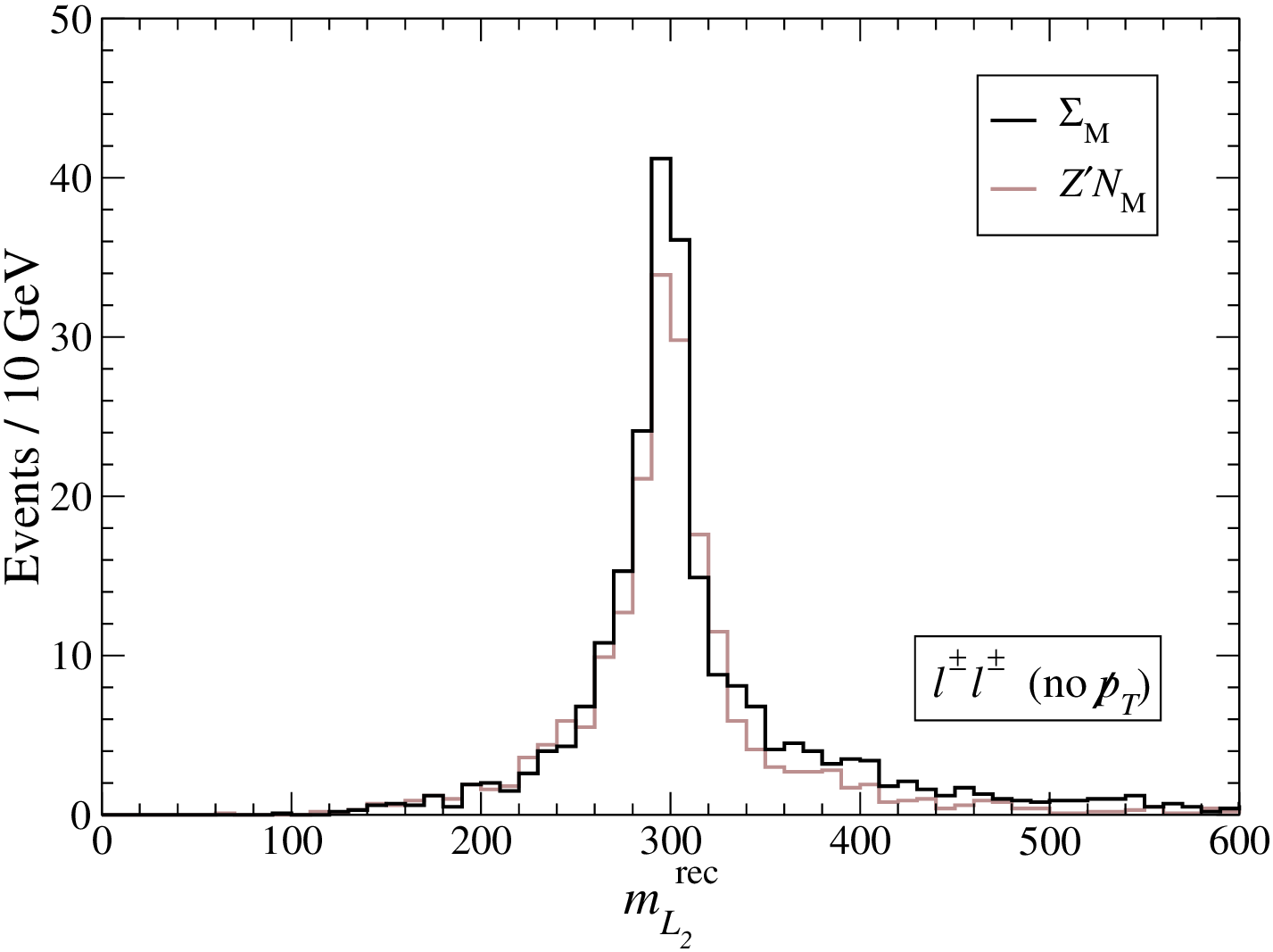,height=5cm,clip=} & \quad &
\epsfig{file=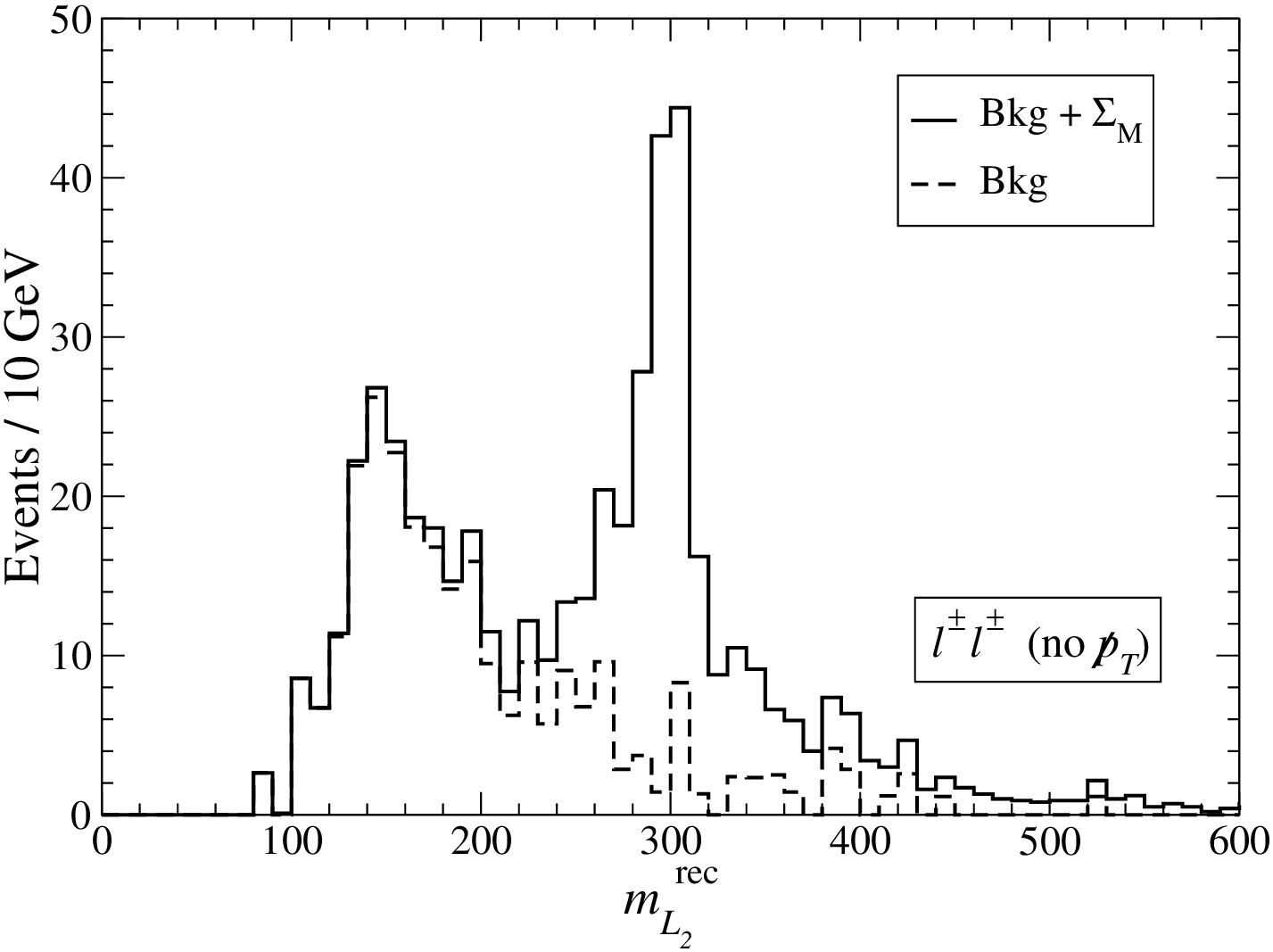,height=5cm,clip=}
\end{tabular}
\caption{Left: reconstructed heavy lepton masses for the signals
in the $\ell^\pm \ell^\pm$ (no $\ptmiss$) sample. Right: the same for the SM background and the background plus the Majorana triplet signal. The luminosity is 30 fb$^{-1}$.}
\label{fig:mrec-2Q2}
\end{center}
\end{figure}

In order to keep only the LNV signals characteristic of a Majorana fermion, we ask for event selection (i) missing energy $\ptmiss < 30$ GeV, and (ii) the presence of at least four hard jets with $p_T > 20$ GeV. In Table~\ref{tab:nsnb-2Q2} we collect the number of signal and background events at pre-selection and selection. We observe that our selection cuts very efficiently select only the signals with LNV neutrino decays, rejecting for example a 97.5\% of the large Dirac triplet signal.
\begin{table}[t]
\begin{center}
\begin{tabular}{cccccccccc}
                      & Pre. & Sel.   & Peak  & \quad &              & Pre. & Sel.   & Peak \\[1mm]
$E^+ E^-$ ($\SM$)     & 21.7  & 1.6   & 0.3   & & $E^+ E^-$ ($\ENd$) & 10.5  & 1.2   & 0.3 \\
$E^\pm N$ ($\SM$)     & 658.0 & 240.0 & 144.8 & & $E^\pm N$ ($\ENd$) & 111.8 & 6.2   & 1.9 \\
$E_1^+ E_1^-$ ($\SD$) & 25.6  & 4.2   & 0.7   & & $N N$ ($\ENd$)     & 47.7  & 1.9   & 0.8 \\
$E_2^+ E_2^-$ ($\SD$) & 0.0   & 0.0   & 0.0   & & $E^+ E^-$ ($\Es$)  & 2.5   & 0.0   & 0.0 \\
$E_1^\pm N$ ($\SD$)   & 174.4 & 9.4   & 2.7   & & $N N$ ($\NM$)      & 433.5 & 202.1 & 132.0 \\
$E_2^\pm N$ ($\SD$)   & 472.0 & 2.9   & 0.9   & & $N N$ ($\ND$)      & 206.0 & 8.1   & 3.1  \\
\hline
$t \bar t nj$         & 1412  & 194   & 7     & & $WW nj$           & 245    & 15    & 3\\
$tW$                  & 96    & 6     & 0     & & $WZnj$            & 1056   & 24    & 1 \\
$W t \bar t nj$       & 184   & 12    & 1     & & $ZZnj$            & 110    & 7     & 1  \\
\end{tabular}
\end{center}
\caption{Number of events in the $\ell^\pm \ell^\pm$ (no $\ptmiss$) sample for the signals and main backgrounds with a luminosity of 30 fb$^{-1}$.}
\label{tab:nsnb-2Q2}
\end{table}
The event reconstruction is performed as follows~\cite{delAguila:2008cj}:
\begin{enumerate}
\item We associate each charged lepton to a pair of jets in all possible ways, using the four jets with larger $p_T$.
\item Among the six possibilities, we choose the one minimising the difference between the two $jj$ and the two $\ell jj$ invariant masses,
\begin{equation}
(m_{j_1 j_2} - m_{j_3 j_4})^2 + (m_{\ell_1 j_1 j_2} - m_{\ell_2 j_3 j_4})^2 \,.
\label{ec:rechad}
\end{equation}
Note that for the leading signal contributions two of the jets in principle correspond to a hadronic $W$ decay and the other two to a $Z$ or Higgs boson decay. However, if a wrong assignment is made, it is expected that the invariant mass differences will be larger.
\end{enumerate}
We present in Fig.~\ref{fig:mrec-2Q2} (left) the reconstructed heavy lepton masses for the two models with heavy Majorana neutrinos which are the only ones yielding observable signals.
On the right side we show the same distributions for the SM background and the background plus the Majorana triplet signal, which has a similar size as the Majorana neutrino singlet one.
The presence of new resonances is apparent and the background normalisation should not be a problem using data away from the peak regions.
We have labelled the resonances as $L_1$, $L_2$ because their identity cannot be established
(their charge is not measured). Actually, for the Majorana triplet signal they are in most cases a charged lepton and a neutral one, while for the $\Zlp$ model both resonances are neutral (see Table~\ref{tab:nsnb-2Q2}). Therefore, the observation of this signal alone points towards the existence of a Majorana (and hence electrically neutral) fermion but its charge is not directly measured as in the trilepton final state.

We define the peak regions as
\begin{align}
250 < m_{L_1}^\text{rec} < 350~\text{GeV} \,, \notag \\
250 < m_{L_2}^\text{rec} < 350~\text{GeV} \,,
\end{align}
and show in Table~\ref{tab:nsnb-2Q2} the number of signal and background events after these cuts.
The statistical significance of the signals for 30 fb$^{-1}$ (neglecting the background uncertainty) is collected in Table \ref{tab:sig-2Q2}, with the luminosity required to achieve $5\sigma$ discovery. Due to the smallness of the background, the discovery criterion is again controlled by the requirement of having at least 10 signal events, and an accurate background normalisation is not crucial because for the discovery luminosities it is very small, around one event.

\begin{table}[ht]
\begin{center}
\begin{tabular}{ccccccc}
       & $\mathcal{S}_0$ & $L$       & \quad &     & $\mathcal{S}_0$ & $L$      \\[1mm]
$\SM$  & 35.0        & 2.1 fb$^{-1}$ & & $\NM$ & 38.3 & 2.3 fb$^{-1}$
\end{tabular}
\end{center}
\caption{Statistical significance for 30 fb$^{-1}$ of the relevant signals, and luminosity $L$ required to have a $5\sigma$ discovery in the $\ell^\pm \ell^\pm$ (no $\ptmiss$) sample.}
\label{tab:sig-2Q2}
\end{table}

\begin{figure}[h]
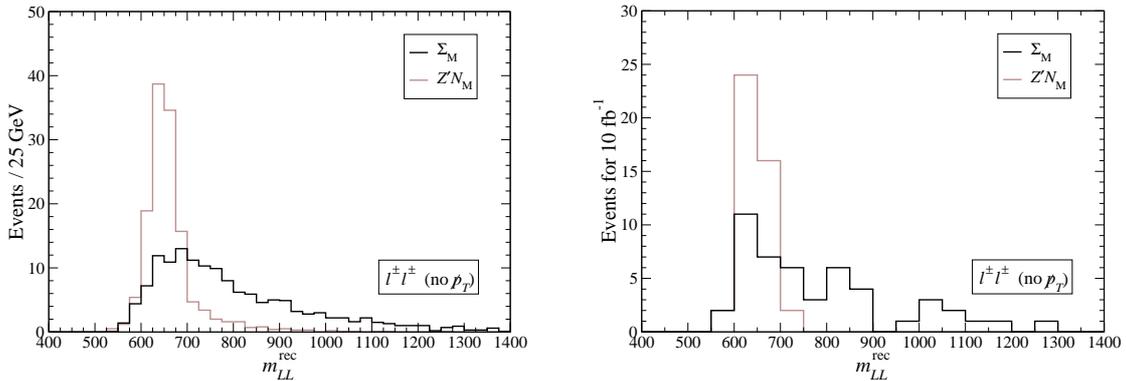

\begin{center}
\begin{tabular}{ccc}
\epsfig{file=Figs/mLL-2Q2-M.eps,height=5cm,clip=} & \quad &
\epsfig{file=Figs/mLL-2Q2-MX.eps,height=5cm,clip=}
\end{tabular}
\caption{Left: invariant mass distribution of the heavy lepton pair in the $\ell^\pm \ell^\pm$ (no $\ptmiss$) sample, after reconstruction cuts. Right: possible experimental result for 10 fb$^{-1}$.}
\label{fig:mLL-2Q2}
\end{center}
\end{figure}

We also address here the discrimination of the two models with Majorana neutrinos in the like-sign dilepton final state. This can easily be done with the heavy lepton pair invariant mass distribution, shown in Fig.~\ref{fig:mLL-2Q2}, which displays a peak for the production by an $s$-channel $Z'$ boson and is more flat for the lepton triplet. In the same figure we show on the right panel a possible experimental result for 10 fb$^{-1}$, obtained with random fluctuations with a Poisson distribution of the bins of the distributions, normalised to the expected total number of events. Both models give quite different results, and the discrimination seems easy once that sufficient luminosity is collected.

\subsection{Final state $\ell^\pm \ell^\pm$ ($\ptmiss$)}
\label{sec:5.2}

\begin{figure}[p]
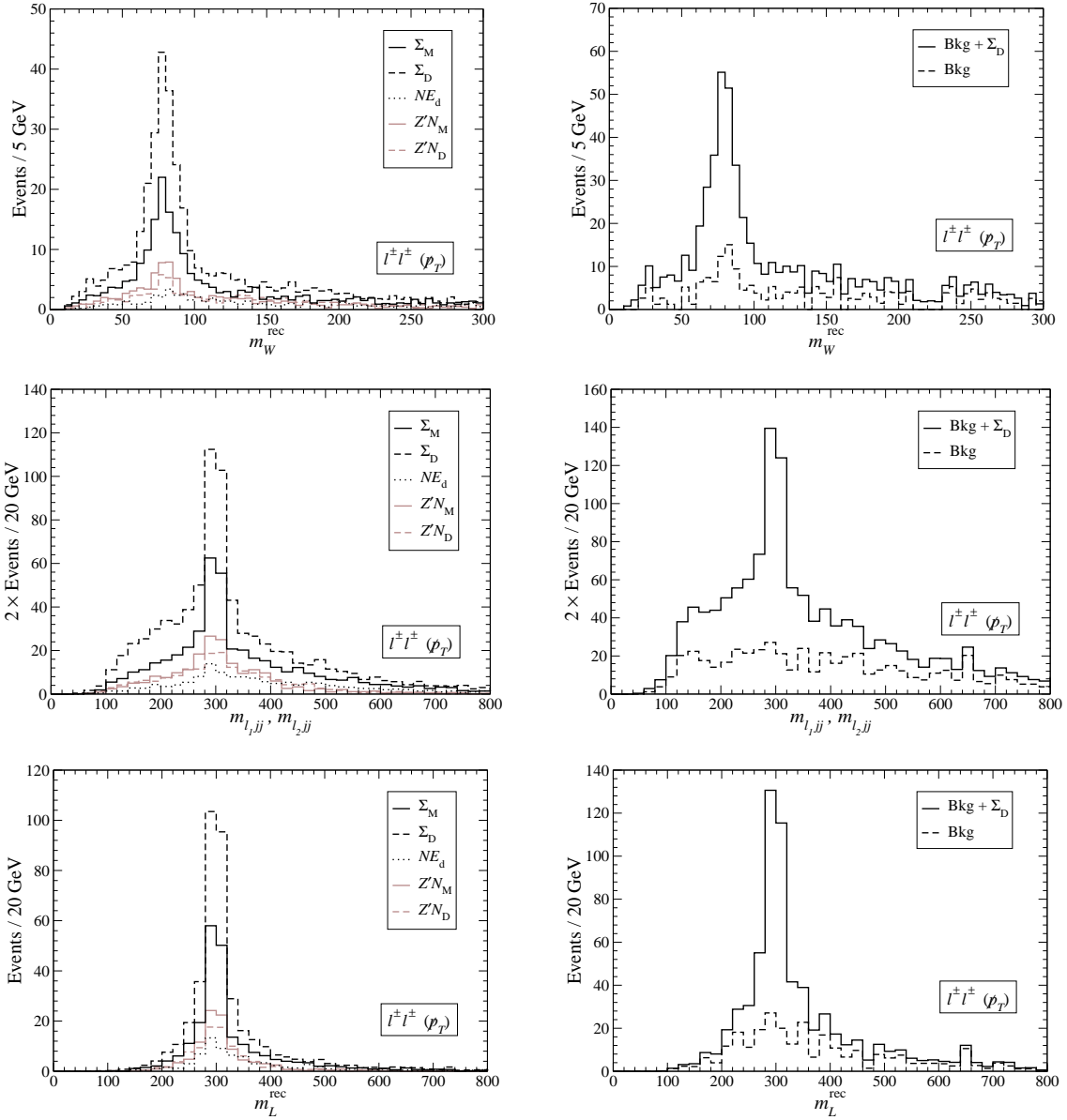

\begin{center}
\begin{tabular}{ccc}
\epsfig{file=Figs/mB-2Q2b.eps,height=5cm,clip=} & \quad &
\epsfig{file=Figs/mB-2Q2b-BS2.eps,height=5cm,clip=} \\[2mm]
\epsfig{file=Figs/mL2x-2Q2b.eps,height=5cm,clip=} & \quad &
\epsfig{file=Figs/mL2x-2Q2b-BS2.eps,height=5cm,clip=} \\[2mm]
\epsfig{file=Figs/mL-2Q2b.eps,height=5cm,clip=} & \quad &
\epsfig{file=Figs/mL-2Q2b-BS2.eps,height=5cm,clip=}
\end{tabular}
\caption{Up: reconstructed $W$ mass for the signals (left) and the Dirac triplet signal with background (right) in the $\ell^\pm \ell^\pm$ ($\ptmiss$) sample.
Middle: the same for the $\ell_1 jj$ and $\ell_2 jj$ invariant mass distribution (two entries per event). 
 Down: the same for the reconstructed heavy lepton mass. The luminosity is 30 fb$^{-1}$.}
\label{fig:mrec-2Q2b}
\end{center}
\end{figure}

The selection criteria to isolate the LNC dilepton signals from Dirac triplets are: (i) large missing energy $\ptmiss > 50$ GeV; (ii) a transverse momentum $p_T > 100$ GeV for the leading charged lepton; (iii) at least two hard jets with $p_T > 20$ GeV. The number of signal and background events is gathered in Table~\ref{tab:nsnb-2Q2b}. Notice that the Majorana triplet also produces a large signal, from a decay similar to the one in Eq.~(\ref{ec:ch2Q2b}) but with lepton number violation (in this model $E^+$ is an anti-fermion).
\begin{table}[htb]
\begin{center}
\begin{tabular}{cccccccccc}
                      & Pre.  & Sel.  & Peak & \quad &               & Pre. & Sel.   & Peak \\[1mm]
$E^+ E^-$ ($\SM$)     & 21.7  & 11.9  & 1.8   & & $E^+ E^-$ ($\ENd$) & 10.5  & 4.1   & 0.3  \\
$E^\pm N$ ($\SM$)     & 658.0 & 215.7 & 84.9  & & $E^\pm N$ ($\ENd$) & 111.8 & 49.2  & 6.8  \\
$E_1^+ E_1^-$ ($\SD$) & 25.6  & 10.0  & 0.7   & & $N N$ ($\ENd$)     & 47.7  & 23.9  & 5.4  \\
$E_2^+ E_2^-$ ($\SD$) & 0.0   & 0.0   & 0.0   & & $E^+ E^-$ ($\Es$)  & 2.5   & 1.5   & 0.5 \\
$E_1^\pm N$ ($\SD$)   & 174.4 & 90.4  & 14.8  & & $N N$ ($\NM$)      & 433.5 & 100.6 & 34.1 \\
$E_2^\pm N$ ($\SD$)   & 472.0 & 301.5 & 155.9 & & $N N$ ($\ND$)      & 206.0 & 90.4  & 22.0  \\
\hline
$t \bar t nj$         & 1412  & 56    & 3     & & $WW nj$           & 245    & 72    & 6 \\
$tW$                  & 96    & 9     & 0     & & $WZnj$            & 1056   & 72    & 1 \\
$W t \bar t nj$       & 184   & 55    & 1     & & $ZZnj$            & 110    & 0     & 0  \\
\end{tabular}
\end{center}
\caption{Number of events in the $\ell^\pm \ell^\pm$ ($\ptmiss$) sample for the signals and main backgrounds with a luminosity of 30 fb$^{-1}$.}
\label{tab:nsnb-2Q2b}
\end{table}
The triplet signals are reconstructed as follows:
\begin{enumerate}
\item  We first reconstruct the $W$ mass as the invariant mass
of the two jets with largest transverse momentum. The result is shown in Fig.~\ref{fig:mrec-2Q2b} (up).
\item  There are two possible pairings with the two charged leptons, $\ell_1 jj$ and $\ell_2 jj$, to reconstruct the heavy lepton invariant mass. We use both to construct a plot with two entries per event
shown in the same figure (middle),
which clearly displays a peak at the true mass for the triplet signals, and small peaks also for the rest. This plot can be used to determine the resonance mass.
\item The reconstruction is completed selecting among the two possibilities the one giving a reconstructed mass close to the value determined from the peak in the previous distributions.
This choice introduces some bias in the non-triplet signals and the SM background, as seen in Fig.~\ref{fig:mrec-2Q2b} (down) but keeps the signal larger.
\end{enumerate}
Notice that despite the appearance of small peaks in the non-triplet signals the differences with the lepton triplet are huge.
We perform kinematical cuts around the peaks,
\begin{align}
60 < m_{W}^\text{rec} < 100~\text{GeV} \,, \notag \\
260 < m_{L}^\text{rec} < 340~\text{GeV} \,,
\end{align}
which are sufficient to reduce the background. The corresponding numbers of events are given in Table~\ref{tab:nsnb-2Q2b}. The statistical significance of the signals, ignoring the background uncertainty, is given in Table~\ref{tab:sig-2Q2b}, with the luminosity required to have a $5\sigma$ significance.
For the Majorana and Dirac triplet signals, which are quite large and observable already with very small luminosity, the background normalisation is not crucial. In the rest of models the figures in Table~\ref{tab:sig-2Q2b} are optimistic but this even enforces our argument, which is to point out that
lepton triplet signals can be seen in this final state while for the rest of models the signals are much harder to observe.
\begin{table}[htb]
\begin{center}
\begin{tabular}{ccccccc}
       & $\mathcal{S}_0$ & $L$     & \quad &     & $\mathcal{S}_0$ & $L$      \\[1mm]
$\SM$  & 23.3        &  3.5 fb$^{-1}$ & & $\NM$ & 9.2 & 13 fb$^{-1}$ \\
$\SD$  & 46.1        &  1.8 fb$^{-1}$ & & $\ND$ & 5.9 & 22 fb$^{-1}$ \\
$\ENd$ & 3.4         &  66 fb$^{-1}$  & & 
\end{tabular}
\end{center}
\caption{Statistical significance for 30 fb$^{-1}$ of the relevant signals, and luminosity $L$ required to have a $5\sigma$ discovery in the $\ell^\pm \ell^\pm$ ($\ptmiss$) sample.}
\label{tab:sig-2Q2b}
\end{table}


\section{Final state $\ell^+ \ell^+ \ell^- \ell^-$}
\label{sec:6}

This four-lepton signal can be produced in several decay channels of $E^+ E^-$ and $E^\pm N$ pairs, both in the case of lepton doublets and triplets, for example
\begin{align}
& E^+ E^- \to \ell^+ Z \, \ell^- Z  \,,
  && \quad ZZ \to \ell^+ \ell^- q \bar q / \nu \bar \nu \,, \nonumber \\ 
& E^+ E^- \to \ell^+ Z \, \ell^- H / \ell^+ H \, \ell^- Z \,,
  && \quad Z \to \ell^+ \ell^- , H \to q \bar q \,, \nonumber \\
& E^+ E^- \to \nu W^+ \ell^- Z / \ell^+ Z \nu W^- \,,
  && \quad Z \to \ell^+ \ell^- , W \to \ell \nu \,, \nonumber \\
& E^\pm N \to \ell^\pm Z \, \ell^- W^+ \,,
  && \quad Z \to \ell^+ \ell^- , W \to q \bar q' \,.
\label{ec:ch4Q0}
\end{align}
For a charged singlet the signal may also result from the $E^+ E^-$ decays indicated, but the cross section is much smaller. This signal is much cleaner than the two previous ones, and our event selection criteria much looser. For pre-selection we choose events having four charged leptons with zero total charge, two of them with transverse momentum $p_T > 30$ GeV and the remaining ones with $p_T > 10$ GeV. For selection we require that opposite-charge leptons cannot be paired in such a way that both pairs have a mass closer to $M_Z$ than 5 GeV. This latter condition is obviously included to reduce the $ZZnj$ background, but hardly affects the signals nor the rest of backgrounds.
\begin{table}[t]
\begin{center}
\begin{tabular}{cccccccccc}
                      & Pres. & Sel.  & Peak  & \quad &              & Pres. & Sel.  & Peak \\[1mm]
$E^+ E^-$ ($\SM$)     & 37.1  & 36.7  & 29.8  & & $E^+ E^-$ ($\ENd$) & 41.6  & 41.2  & 34.2 \\
$E^\pm N$ ($\SM$)     & 26.2  & 25.8  & 16.0  & & $E^\pm N$ ($\ENd$) & 79.3  & 78.6  & 54.8 \\
$E_1^+ E_1^-$ ($\SD$) & 137.6 & 135.8 & 111.4 & & $N N$ ($\ENd$)     & 38.2  & 38.2  & 12.0 \\
$E_2^+ E_2^-$ ($\SD$) & 0.0   & 0.0   & 0.0   & & $E^+ E^-$ ($\Es$)  & 5.2   & 5.1   & 4.0 \\
$E_1^\pm N$ ($\SD$)   & 83.5  & 82.6  & 56.8  & & $N N$ ($\NM$)      & 62.3  & 62.0  & (18.1) \\
$E_2^\pm N$ ($\SD$)   & 0.0   & 0.0   & 0.0   & & $N N$ ($\ND$)      & 130.7 & 128.9 & (41.5) \\
\hline
$t \bar t nj$         & 30    & 30    & 1     & & $Z t \bar t nj$    & 20    & 20    & 2 \\
$Z b \bar b nj$       & 11    & 11    & 0     & & $ZZnj$             & 599   & 129   & 7 \\
\end{tabular}
\end{center}
\caption{Number of $\ell^+ \ell^+ \ell^- \ell^-$ events for the signals and main backgrounds with a luminosity of 30 fb$^{-1}$.}
\label{tab:nsnb-4Q0}
\end{table}
The number of signal and background events at the two stages of event selection is collected in Table~\ref{tab:nsnb-4Q0}.
The event reconstruction is performed following the procedure in Ref.~\cite{delAguila:2008cj}, analogous to the one used for $\ell^\pm \ell^\pm$ signals with missing energy:
\begin{enumerate}
\item First, the two charged leptons coming from the $Z$ boson decay are identified, selecting among the three possibilities the opposite sign pair $\ell_a^+ \ell_b^-$ which has an invariant mass closest to $M_Z$.
\item Then, the presence of a heavy charged lepton is investigated using a plot with two entries per event,
corresponding to the invariant mass of the $Z$ candidate plus one of the remaining charged leptons, $m(\ell_a^+ \ell_b^- \ell_c)$ and $m(\ell_a^+ \ell_b^- \ell_d)$. 
\item Once that the location of the peak is found, we can determine which leptons are the $E$ decay products by choosing between the two possibilities $\ell_a^+ \ell_b^- \ell_c$ and $\ell_a^+ \ell_b^- \ell_d$, the one giving an invariant mass closest to $m_E$. The reconstructed $E$ mass $m_E^\text{rec}$ is then the three-lepton invariant mass.
\end{enumerate}
\begin{figure}[t]
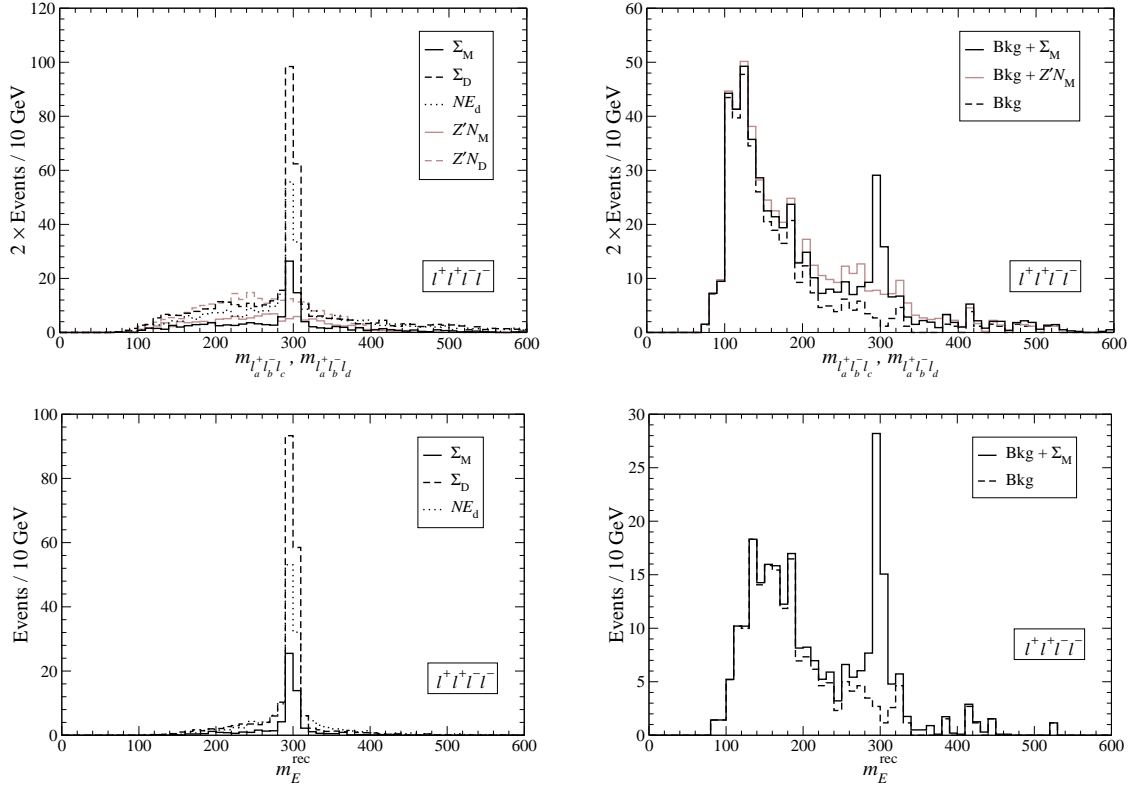

\begin{center}
\begin{tabular}{ccc}
\epsfig{file=Figs/mE2x-4Q0.eps,height=5cm,clip=} & \quad &
\epsfig{file=Figs/mE2x-4Q0-BS15.eps,height=5cm,clip=} \\[2mm]
\epsfig{file=Figs/mE-4Q0.eps,height=5cm,clip=} & \quad &
\epsfig{file=Figs/mE-4Q0-BS1.eps,height=5cm,clip=} 
\end{tabular}
\caption{Up, left: $\ell_a^+ \ell_b^- \ell_c$, $\ell_a^+ \ell_b^- \ell_d$ invariant mass distribution for the five significant signals in the $\ell^+ \ell^- \ell^+ \ell^-$ final state (with two entries per event). Up, right: the same for the SM background and the background plus two of the signals. Down, left: reconstructed heavy charged lepton masses for several of the signals in this final state. Down, right: the same for the SM background and the background plus the Majorana triplet signal. The luminosity is 30 fb$^{-1}$.}
\label{fig:mrec-4Q0}
\end{center}
\end{figure}
We show in Fig.~\ref{fig:mrec-4Q0} how this procedure would work. In the upper left panel we show
the $\ell_a^+ \ell_b^- \ell_c$, $\ell_a^+ \ell_b^- \ell_d$ invariant mass distribution for the five sizeable signals. Those with a heavy charged lepton exhibit a peak, as expected, while for the rest the distributions are broad. 
On the upper right panel we show the same distribution for the background and also including two of the signals: the Majorana triplet (which has a resonance) and the Majorana singlet (which does not) to illustrate the difference. In the latter models, a four-lepton excess would be detected over the background but not corresponding to a charged resonance.
Once that a peak is found, the $E$ mass can be reconstructed, as shown in Fig.~\ref{fig:mrec-4Q0} (down, left) for the signals which display a peak.
It is important to note that the selection between the two reconstructed mass values does not significantly bias the background, as seen in the lower right panel.
We take the peak region as the interval
\begin{align}
280 < m_E^\text{rec} < 320~\text{GeV} \,,
\end{align}
and perform a kinematical cut to find the signal significances. The number of events at the peak are given in Table~\ref{tab:nsnb-4Q0}. For completeness, we give in parentheses the number of events for the non-resonant signals. The statistical significance (neglecting the background uncertainty) and the luminosity required for a $5\sigma$ discovery are given in Table~\ref{tab:sig-4Q0}.

\begin{table}[ht]
\begin{center}
\begin{tabular}{ccccccc}
       & $\mathcal{S}_0$ & $L$       & \quad &     & $\mathcal{S}_0$ & $L$      \\[1mm]
$\SM$  & 14.2        & 6.6 fb$^{-1}$ & & $\ENd$ & 31.3 & 3.0 fb$^{-1}$ \\
$\SD$  & 52.2        & 1.8 fb$^{-1}$ & & 
\end{tabular}
\end{center}
\caption{Statistical significance for 30 fb$^{-1}$ of the relevant signals, and luminosity $L$ required to have a $5\sigma$ discovery in the $\ell^+ \ell^- \ell^+ \ell^-$ final state.}
\label{tab:sig-4Q0}
\end{table}

\section{Summary: model discrimination}
\label{sec:7}

The results in the previous sections can be summarised in Table~\ref{tab:summ}, which collects the luminosities required for $5\sigma$ discovery for each model and decay channel. The trilepton final state is split into the sample without $Z$ candidates (no $Z$) and with a $Z$ candidate $(Z)$, and the like sign dilepton into samples without and with missing energy. The data in this table make apparent that the different models considered give different signals, and the complementarity of the channels studied make it possible to identify the nature of a signal eventually observed at LHC. 

\begin{table}[htb]
\begin{center}
\begin{tabular}{cccccc}
      & $\ell^\pm \ell^\pm \ell^\mp$ (no $Z$) & $\ell^\pm \ell^\pm \ell^\mp$ ($Z$)  & $\ell^\pm \ell^\pm$ (no $\ptmiss$) & $\ell^\pm \ell^\pm$ ($\ptmiss$) & $\ell^+ \ell^+ \ell^- \ell^-$ \\
$\SM$  & 3.3   & 25  & 2.1   & 3.5  & 6.6 \\
$\SD$  & 1.5   & 17  & --    & 1.8  & 1.8 \\
$\ENd$ & 1.1   & --  & --    & --   & 3.0 \\
$\Es$  & --    & --  & --    & --   & --  \\
$\NM$  & 2.1 P & --  & 2.3 P & 13   & --  \\
$\ND$  & 1.1 P & --  & --    & 22   & -- \\
\end{tabular}
\end{center}
\caption{Luminosities (in fb$^{-1}$) required for $5\sigma$ discovery for the models in the left column in the final states indicated. The presence of a peak in the heavy lepton pair invariant mass is indicated with a ``P''. A dash indicates an unobservable signal, or a discovery luminosity larger than 30 fb$^{-1}$.}
\label{tab:summ}
\end{table}

All the models except the charged lepton singlet (which will be commented below) give signals in the
$\ell^\pm \ell^\pm \ell^\mp$ (no $Z$) final state, and searches in this channel are very sensitive to the presence of a heavy Dirac or Majorana neutrino in a singlet, doublet or triplet representation. For an integrated luminosity of 100 fb$^{-1}$, the approximate mass reach of this final state 
for the different models with heavy neutrinos is of 675 GeV ($\SM$), 800 GeV ($\SD$), 850 GeV ($\ENd$), 850 GeV ($\NM$) and 1 TeV ($\ND$) for $5\sigma$ evidence.
Model discrimination is possible with the analysis of the remaining signals:
\begin{enumerate}
\item Dirac / Majorana: the presence of like-sign dilepton signals without missing energy indicates the Majorana nature of the neutrino (models $\SM$, $\NM$) and its absence its Dirac nature (models $\SD$, $\ENd$ or $\ND$).
\item Majorana triplet ($\SM$) / Majorana singlet ($\NM$): these can be easily distinguished because the former gives
(i) four-lepton signals;  
(ii) much larger like-sign dilepton signals with missing energy; and
(iii) a signal (with a peak at $m_E$) in the $\ell^\pm \ell^\pm \ell^\mp$ ($Z$) final state,
while the latter gives (iv) a peak in the heavy lepton pair invariant mass distribution, in the $\ell^\pm \ell^\pm \ell^\mp$ (no $Z$) and $\ell^\pm \ell^\pm$ final states.
\item Dirac singlet ($\ND$) / lepton doublet and Dirac triplet ($\ENd$, $\SD$): the Dirac singlet
(i) does not give four lepton signals as the others;
(ii) it has a peak in the heavy lepton pair invariant mass distribution.
\item Lepton doublet ($\ENd$) / Dirac triplet ($\SD$): these the most alike signals, but they can nevertheless be distinguished by the presence in the case of the triplet of
(i) a large like-sign dilepton signal with missing energy;
(ii) much larger four lepton signals, as large as the trilepton ones; and
(iii) a signal in the $\ell^\pm \ell^\pm \ell^\mp$ ($Z$) final state.
\end{enumerate}
Finally, we comment on the charged singlet model. The signals in this case are unobservable due to their small cross section. (Signals in the opposite-charge dilepton channel are difficult to see
even for the lepton triplet because of the large backgrounds~\cite{delAguila:2008cj,delAguila:2008hw}, and for the charged singlet they are even smaller than in those models.)
However, in a model with a $Z'$ boson coupling to them, charged singlet production would be enhanced as it is for neutrino singlets. In such case, their presence  would be characterised by: (i) four lepton signals with a reconstructable peak at $m_E$; (ii) absence of $\ell^\pm \ell^\pm \ell^\mp$ (no $Z$) signals or small ones; (iii) absence of like-sign dilepton signals.

\section{Conclusions}
\label{sec:8}

In order to be adequate for new physics searches, a final state signature must have two properties: first, it has to be sensitive to new physics, {\em i.e.} to be a possible signal produced in models beyond the SM; second, it must have a small background, at least compared to the signals expected. The trilepton final state
$\ell^\pm \ell^\pm \ell^\mp$ shares both features for heavy neutrino searches at LHC: it is sensitive to
Majorana or Dirac neutrinos in triplet, doublet or singlet $\text{SU}(2)_L$ representations and it has a small SM background. The broad sensitivity of this final state, apparent with a glance at Table~\ref{tab:summ}, is unique because other clean signals,
as for example like-sign dileptons without missing energy, are produced by heavy Majorana neutrinos but not by Dirac neutrinos. Moreover, the trilepton final state is the only one in which the presence of a neutral heavy particle (reconstructed as a peak in the invariant mass of two opposite-charge leptons plus missing energy) can be established. And, with a higher luminosity, it can establish the triplet nature of the heavy neutrino.
On the other hand, the trilepton SM background can be practically removed with adequate event selection criteria. For these reasons, the trilepton final state can properly be considered as the golden channel for heavy neutrino searches at LHC.

The like-sign dilepton final state is essential to elucidate the Majorana or Dirac character of a heavy neutrino. Large $\ell^\pm \ell^\pm$ signals are produced in several of the models studied, not only from LNV processes (as in the case of a heavy Majorana neutrino) but also from LNC ones (as for a Dirac triplet), being the main difference the presence of final state neutrinos in the latter case. Therefore, we have performed two different analyses for the like-sign dilepton final state. In the first one our selection criteria, in particular the absence of significant missing energy, suppress the like-sign dilepton signals produced in LNC processes and efficiently isolate the true LNV ones, so that an eventually observed signal would correspond to a Majorana neutrino and its absence would indicate its Dirac character. In the second one we have required the opposite: large missing energy. With this novel analysis we have found that Dirac triplets give large and observable $\ell^\pm \ell^\pm$ signals with large missing energy, while doublets and singlets do not. This is a notable result, because like-sign dileptons are usually regarded as a characteristic signature of Majorana, not Dirac fermions. (Of course, the difference in our analysis is the requirement of large missing energy, otherwise LNC signals would be suppressed.) This dilepton signal is very distinctive, and could be used to discriminate Dirac triplets from the other models with a heavy Dirac neutrino.

The third final state examined, $\ell^+ \ell^+ \ell^- \ell^-$, proves the presence of a heavy charged lepton $E$, with the observation of a sharp peak in a trilepton invariant mass distribution. Moreover, it can easily determine the triplet, doublet or singlet character of the neutrino, combined with the information from the like-sign dilepton final state. The charged singlet would also give a clean signal in this sample if its production is enhanced, {\em e.g.} by the presence of a $Z'$ boson. The information of this four lepton channel is complemented by other indications, such as the
observation of a peak at $m_E$ in the trilepton final state with a $Z$ candidate, or of a resonance in the heavy lepton pair invariant mass distribution, which help identify the underlying model with a luminosity not much larger than the one required for $5\sigma$ discovery. Additional final states studied in Refs.~\cite{delAguila:2008cj,delAguila:2008hw}, for example $\ell^\pm \ell^\pm \ell^\pm \ell^\mp$
and $\ell^\pm \ell^\pm \ell^\pm$, would provide further information but only at much larger integrated luminosities.

In conclusion, in this paper we have shown that, if a positive signal of new heavy leptons is found, the six models considered can be discriminated already at LHC with the study of multi-lepton signals, without the need of heavy lepton coupling measurements at future colliders.
An extra bonus of our comparative analysis is that it provides a guide of consistency checks to be performed (and final states to be investigated) in case that a multi-lepton event excess compatible with heavy lepton pair production is found at LHC. We have assumed the Higgs boson to be light, as preferred by electroweak precision data, taking a mass of 115 GeV. For a heavier Higgs the results are expected to be similar although a dedicated analysis is required.
Our study has been done at the level of a fast detector simulation of all the signal contributions (which is a non-trivial task) and the relevant SM backgrounds. The former is important, because there are many possible decay channels for the heavy leptons, and usually several of them contribute to a final state with a given charged lepton multiplicity. The latter is important as well. Although our simulation is not as detailed as one with a complete detector description, it incorporates some of the features found in real experiments, which parton-level analyses fail to correctly estimate. But of course, a full detector simulation is desirable, and it will have to be performed in order to compare the predictions for new lepton signals with the forthcoming LHC data.

\section*{Acknowledgements}

I thank F. del Aguila and M. P\'erez-Victoria for useful comments.
This work has been supported by a MEC Ram\'on y Cajal contract, MEC project FPA2006-05294 and
Junta de Andaluc{\'\i}a projects FQM 101 and FQM 437.

\appendix
\section{Partial widths for heavy lepton decays}
\label{sec:a}

We give here the partial widths for heavy lepton decays in the different models considered, omitting the vanishing ones. For the Majorana triplet we have
\begin{align}
& \Gamma(E^+ \to \bar \nu W^+) = \frac{g^2}{32 \pi} |V_{lN}|^2
\frac{m_E^3}{M_W^2} \left( 1- \frac{M_W^2}{m_E^2} \right) 
\left( 1 + \frac{M_W^2}{m_E^2} - 2 \frac{M_W^4}{m_E^4} \right) \equiv \Gamma_{\nu W} \,, \nonumber
\\[0.1cm] 
& \Gamma(E^+ \to l^+ Z) =  \frac{g^2}{64 \pi c_W^2} |V_{lN}|^2
\frac{m_E^3}{M_Z^2} \left( 1- \frac{M_Z^2}{m_E^2} \right) 
\left( 1 + \frac{M_Z^2}{m_E^2} - 2 \frac{M_Z^4}{m_E^4} \right) \equiv \Gamma_{lZ} \,, \nonumber
\\[0.1cm] 
& \Gamma(E^+ \to l^+ H) =  \frac{g^2}{64 \pi} |V_{lN}|^2
\frac{m_E^3}{M_W^2} \left( 1- \frac{M_H^2}{m_E^2} \right)^2 \equiv \Gamma_{l H} \,, \notag \\
& \Gamma(N \to l^- W^+) = \frac{g^2}{64 \pi} |V_{l N}|^2
\frac{m_N^3}{M_W^2} \left( 1- \frac{M_W^2}{m_N^2} \right) 
\left( 1 + \frac{M_W^2}{m_N^2} - 2 \frac{M_W^4}{m_N^4} \right)
\equiv \Gamma_{lW} \,, \nonumber \\[0.1cm] 
\displaybreak
& \Gamma(N \to l^+ W^-) = \Gamma_{lW} \,, \notag \\ 
& \Gamma(N \to \nu_l Z) =  \frac{g^2}{64 \pi c_W^2} |V_{lN}|^2
\frac{m_N^3}{M_Z^2} \left( 1- \frac{M_Z^2}{m_N^2} \right) 
\left( 1 + \frac{M_Z^2}{m_N^2} - 2 \frac{M_Z^4}{m_N^4} \right)
\equiv \Gamma_{\nu Z} \,, \nonumber \\[0.2cm]
& \Gamma(N \to \nu_l H) =  \frac{g^2}{64 \pi} |V_{lN}|^2
\frac{m_N^3}{M_W^2} \left( 1- \frac{M_H^2}{m_N^2} \right)^2
\equiv \Gamma_{\nu H} \,.
\label{ec:Nwidths}
\end{align}
For the Dirac triplet,
\begin{align}
& \Gamma(E_2^+ \to \nu_l W^+) = \Gamma_{\nu W} \,, && \Gamma(N \to l^- W^+) = \Gamma_{lW} \,, \notag \\
& \Gamma(E_1^- \to l^- Z) = \Gamma_{lZ} \,, && \Gamma(N \to \nu_l Z) = \frac{1}{2} \Gamma_{\nu Z} \,, \notag \\
& \Gamma(E_1^- \to l^- H) = \Gamma_{lH} \,, && \Gamma(N \to \nu_l H) = \frac{1}{2} \Gamma_{\nu H}  \,.
\end{align}
For the isodoublet the only non-vanishing ones are
\begin{align}
& \Gamma(E^+ \to l^+ Z) = \frac{1}{2} \Gamma_{lZ} \,, && \Gamma(N \to l^- W^+) = \Gamma_{lW} \,, \notag \\
& \Gamma(E^+ \to l^+ H) = \frac{1}{2} \Gamma_{lH} \,,
\end{align}
while for the isosinglet $E$ we have
\begin{align}
& \Gamma(E^+ \to \nu_l W^+) = \frac{1}{2} \Gamma_{\nu W} \,,  \notag \\
& \Gamma(E^+ \to l^+ Z) = \frac{1}{2} \Gamma_{lZ} \,, \notag \\
& \Gamma(E^+ \to l^+ H) = \frac{1}{2} \Gamma_{lH} \,.
\end{align}
The widths for a heavy Majorana neutrino singlet are the same as for the triplet,
\begin{align}
& \Gamma(N \to l^- W^+) = \Gamma_{lW} \,, \notag \\
& \Gamma(N \to l^+ W^-) = \Gamma_{lW} \,, \notag \\
& \Gamma(N \to \nu_l Z) = \Gamma_{\nu Z} \,, \notag \\
& \Gamma(N \to \nu_l H) = \Gamma_{\nu H}  \,,
\end{align}
and for a Dirac neutrino singlet they are
\begin{align}
& \Gamma(N \to l^- W^+) = \Gamma_{lW} \,, \notag \\
& \Gamma(N \to \nu_l Z) = \frac{1}{2} \Gamma_{\nu Z} \,, \notag \\
& \Gamma(N \to \nu_l H) = \frac{1}{2} \Gamma_{\nu H}  \,.
\end{align}

\end{document}